\documentclass{aa}
\usepackage{graphicx}
\usepackage{soul}
\usepackage{txfonts}
\pdfoutput=1
\usepackage{color}

\begin{document} 

\title{The Lyman continuum escape fraction and the Mean Free Path of hydrogen ionizing photons for bright
$z\sim 4$ QSOs from SDSS DR14}

\author{M. Romano\inst{1}
\and A. Grazian\inst{2,3}
\and E. Giallongo\inst{3}
\and S. Cristiani\inst{4}
\and F. Fontanot\inst{4}
\and K. Boutsia\inst{5}
\and F. Fiore\inst{4}
\and N. Menci\inst{3}
}

\offprints{M. Romano, \email{michael.romano@studenti.unipd.it}}

\institute{Dipartimento di Fisica e Astronomia, Universit\`a di Padova,
Vicolo dell'Osservatorio 3, I-35122, Padova, Italy
\and
INAF--Osservatorio Astronomico di Padova, Vicolo dell'Osservatorio 5, I-35122, Padova, Italy
\and
INAF--Osservatorio Astronomico di Roma, Via Frascati 33,
I-00078, Monte Porzio Catone, Italy
\and
INAF--Osservatorio Astronomico di Trieste, Via G.B. Tiepolo 11, I-34143, Trieste, Italy
\and
Las Campanas Observatory, Carnegie Observatories, Colina El Pino Casilla 601, La Serena, Chile
}

\date{Received 26 March 2019; Accepted 4 October 2019}

\authorrunning{Romano et al.}
\titlerunning{The LyC $f_{esc}$ and MFP of bright SDSS QSOs at $z\sim 4$}
 
  \abstract
{
One of the major challenges in observational cosmology is related to the redshift evolution of the average hydrogen (HI) ionization in the Universe, as evidenced by the changing in the ionization level of the intergalactic medium (IGM) through cosmic time. In particular, starting from the first cosmic reionization, the rapid evolution of the IGM physical properties poses severe constraints for the identification of the sources responsible for keeping its high level of ionization up to lower redshifts.
}
{
In order to probe the ionization level of the IGM and the ionization capabilities of bright QSOs at $z=4$, we have selected a sample of 2508 QSOs drawn from the SDSS survey (DR14) in the redshift interval $3.6 \leq z \leq 4.6$ and absolute magnitude range $-29.0 \lesssim M_{1450} \lesssim -26.0$. Particularly, we focus on the estimate of the escape fraction of HI ionizing photons and their mean free path (MFP), which are fundamental in the characterization of the surrounding intergalactic medium.
}
{
Starting from UV/optical rest-frame spectra of the whole QSO sample from the SDSS survey, we estimate the escape fraction and free path individually for each of them. We calculate the Lyman Continuum escape fraction as the flux ratio blue-ward ($\sim$ 900 {\AA} rest-frame) and red-ward ($\sim$ 930 {\AA} rest-frame) of the Lyman limit (LL). We then obtain the probability distribution function (PDF) of the individual free paths of the QSOs in the sample and study its evolution in luminosity and redshift, comparing our results with the ones found in literature so far.
}
{
We find a lower limit to the mean Lyman Continuum escape fraction of $0.49$, in agreement with the values obtained for both brighter and fainter sources at the same redshift. We show that the free paths of ionizing photons are characterized by a skewed distribution function peaked at low values, with an average of $\sim49-59$ proper Mpc at $z\sim4$, after excluding possible associated absorbers (AAs). This value is larger than the one obtained at the same redshift by many authors in the literature using different techniques.
Moreover, the probability distribution function of free path gives a complementary information w.r.t. the mean free path derived through the stack technique.
Finally, we also find that the redshift evolution of this parameter results to be possibly milder than previously thought.
}
{
Our new determination of the mean free path at $z\sim 4$ implies that previous estimates of the HI photo-ionization rate $\Gamma_\mathrm{HI}$ available in the literature should be corrected by a factor of 1.2-1.7.
These results have important implications if extrapolated at the epoch of reionization.
}


\maketitle
%

\section{Introduction}

The epoch of reionization (EoR) marks a fundamental event for the
Universe, characterized by a transition phase of the intergalactic
medium (IGM) from cold and almost neutral to warm and fully
ionized \citep{meiksin09,mcquinn16}. During this dramatic event,
located approximately at $z\sim 6-8$, the first sources of UV photons
with energy above 13.6 eV were able to clear the fog of the widespread
neutral hydrogen (HI) and put an end to the so called period of the
Dark Ages \citep{dayal18}. Only in the recent years, thanks mainly to
the analysis of the CMB optical depth by Planck \citep{planck18} and
the high-z QSOs \citep{fan06,becker15}, it was clear that the EoR was a
very rapid event; this phase transition lasted for a short time period
$\Delta z\le 2.8$, and it also was a patchy process. This is consistent
with the progressive decrease of the photo-ionization rate
$\Gamma_\mathrm{HI}$ observed at $z\ge 5.5$ \citep{calverley11,davies18,daloisio18}.

A long-lasting debate is in progress on the sources responsible for
the EoR. While the majority of the astrophysical community are
favoring star-forming galaxies (SFGs) as the main drivers of HI
reionization \citep{dayal18}, alternative explanations based on Active
Galactic Nuclei (AGNs) have been
proposed \citep{giallongo12,giallongo15,madau15,grazian18,boutsia18}.
It is widely accepted that QSOs and AGNs dominate the HI
ionizing background at $z<3$ \citep{hm12,fontanot12,cristiani16,fg19}. At higher
redshifts, however, the debate is still on-going whether the HI ionizing
background is mainly driven by rest-frame star-forming galaxies or by AGNs.
One of the main criticism against HI reionization driven by accreting super massive black holes (SMBHs) is the fast drop of the space density of bright QSOs \citep{fan01,cowie09} at $z\ge 3$ and the lack of numerous $L\sim L^*$ AGNs at $z\ge 4$ \citep{parsa18,akiyama18,kim18,matsuoka18,yang18}. In fact, contrasting results have been obtained in literature so far, leading to significant uncertainties in the estimate of the HI photo-ionization rate, in which the luminosity function directly enters \citep{boutsia18,giallongo19,kulkarni18}. 

In order to quantify the contribution of active SMBHs to the
cosmological photo-ionizing background, two other physical parameters
are needed, in addition to the knowledge of the QSO luminosity function:
1-the escape fraction of HI ionizing photons, i.e.,
at the Lyman continuum (LyC) $\lambda\sim 900$ {\AA}
rest-frame, $f_{esc}(\mathrm{LyC})$, and
2-the mean free path (MFP) of HI ionizing photons. The latter is the average physical
distance HI-ionizing photons typically travel before being
absorbed by a factor $1/e$, i.e., by an optical depth of $\tau_\mathrm{HI}=1$.
This quantity directly enters into the calculation of the photo-ionization
rate $\Gamma_\mathrm{HI}$, and it can thus change in a
significant way the amount of UV photons per unit time released by
QSOs into the IGM.

At $z\sim 1$, \citet{cowie09} find $f_{esc}(\mathrm{LyC})\sim 100\%$
for bright QSOs, while \citet{stevans14} indicate similar results
for a sample of fainter Seyfert 1 and Seyfert 2 galaxies.
The LyC escape fraction of bright QSOs ($M_{1450}\lesssim -26$) and
faint AGNs ($M_{1450}\sim -24$) at $z\sim 4$ has been studied
by \citet{cristiani16} and \citet{grazian18}, respectively. They find
that the Lyman Continuum escape fraction of QSOs and AGNs is $\sim
75\%$ or greater, and it does not show any dependence on the absolute
luminosities of the objects, at least up to $L\sim L^*$.
In particular, \citet{cristiani16} find that the distribution of the
escape fraction of bright QSOs from SDSS is bimodal, with 
18-25\% probability of negligible escape fraction and the remaining 75-82\%
with $f_{esc}(\mathrm{LyC})\sim 100\%$. However, as discussed in \citet{grazian18},
this bi-modality could be due to their choice of the spectral window adopted to
measure the ionizing photon leakage, i.e., $\lambda_{rest}=865-885$ {\AA}.
Indeed, sources with high escape fraction and a proximate Lyman Limit
System (LLS) or a Damped Lyman-$\alpha$ system (DLA), close to the emission
redshift of a QSO, can spuriously mimic an $f_{esc}(\mathrm{LyC})\sim 0$.
It is thus important to study in detail the relation between the LyC escape
fraction and the presence of nearby absorbers. An effective parameter which is useful to quantify the presence of LLSs surrounding QSOs could be the mean free path.

The individual free path (IFP) of HI ionizing photons
can be obtained by deriving the optical depth along a
line of sight by counting the individual intervening absorbers
\citep{fg08,songaila10,rudie13,omeara13,inoue14,prochaska15,crighton18}.
This approach yielded different results depending on the input HI column
density distribution used in the models.
The robust detection of absorbers, and in particular (partial) LLSs however, is a delicate issue and it has been
limited to the line of sight of very luminous QSOs. They can reside in
very massive halos, thus biasing our view of the MFP with respect to the one
in the other, less biased, regions of the Universe.
Alternatively, the mean free path can be obtained by stacking a large number
of high-z QSOs and measuring the decrement of a factor $1/e$ with respect to
the mean flux level at $\lambda_{rest}\sim 912$ {\AA}, as carried out by
\cite{prochaska09,worseck14}. They obtain a fast evolution of the MFP with
the cosmic epoch MFP$(z) \propto (1+z)^{-5.4}$, over a large redshift range
$2.3<z<5.5$.
This rapid decrement implies, according to \citet{worseck14}, an evolution in
number density or on the physical size of the absorbers.

These two methods (individual absorber distribution vs stacking),
however, are not providing fully consistent results.
In particular, the redshift evolution of the MFP derived by \cite{worseck14}
is particularly rapid compared with the one found by the statistics of
the absorbers, (e.g., \cite{songaila10}). For example,
\citet{songaila10} find a number density of LLSs per unit redshift
$n(z)\propto (1+z)^{-1.94}$, corresponding to a mean free path evolution of
MFP$(z)\propto (1+z)^{-4.44}$.
For comparison, \citet{crighton18}  measure a redshift evolution for the incidence of LLS of $l(z)\propto (1 + z)^{1.70}$.
These differences however are not surprising, since different absorber populations (e.g., partial Lyman limit systems, Lyman limit systems, damped systems) could contribute in a different way to the mean free path and scale differently with redshift \citep{prochaska10,inoue14}.

The mean free path of HI ionizing photons is a physical quantity
averaged over a sample of QSOs at high-z. It is a measure of the
distribution of Lyman limit opacity and, hence, of the IGM ionization
state in the Lyman continuum. More information is carried out by the
probability distribution function of the individual free paths of the
population, PDF(IFP). This latter depends on the distribution of
neutral patches, sparse HI residuals surrounded by a widespread, ionized
gas. They are constituted both by diffuse gas clouds in the IGM and/or
by denser large scale structures, i.e., the external regions of galaxies
or around filaments, as proposed by \cite{fumagalli11,worseck14}.

The study of PDF(IFP) is important as a benchmark for numerical
simulations of the IGM. Strong constraints on cosmological simulations
can be provided by a detailed knowledge of the IFP distribution across
a wide redshift range. For example, radiation-hydrodynamical
simulations by \citet{rahmati18} indicate that, close to the
reionization epoch (at $z=6$), the distribution of IFP peaks at
$\sim$5 pMpc with an extended tail towards smaller values, if measured
along a random line of sight. If instead the line of sight is centered
on massive galaxies, they predict a bimodal distribution with a peak
at few kpc and another, less pronounced, peak at $\sim$5 proper Mpc.

In this paper we aim at studying the distribution function of the
HI ionizing individual free paths centered on relatively bright
($M_{1450}\lesssim -26$) QSOs at $3.6 \leq z \leq 4.6$, and we try to understand the
connection between the IFP and the LyC escape fraction.
This work can be useful to understand the physical conditions that
allow HI ionizing photons to escape from QSOs and subsequently
to ionize the IGM.

The paper is organized as follows: in Section 2 we describe
the SDSS data used in this work, in Section 3 we outline the adopted method to measure the QSO physical properties (i.e., spectral slope, Lyman Continuum escape fraction, individual free path of HI ionizing photons). Section 4 describes the
results, Section 5 investigates the connection between $f_{esc}(\mathrm{LyC})$ and free path. Discussions, summary, and conclusions are provided in Sections 6 and 7, respectively. In Appendix A we provide an example of SDSS QSO spectrum, while in Appendix B we
discuss the difference between the MFP obtained from stacked spectra and from the mean value of the IFP distribution.

Throughout the paper, we assume H$_0=70$ km/s/Mpc, $\Omega_m=0.3$, and
$\Omega_\Lambda=0.7$. Physical distances are expressed in proper Mpc (pMpc).


\section{Data}

\subsection{The SDSS DR14 QSO sample}

The 14th Data Release of the Sloan Digital Sky Survey (SDSS, DR14) contains observations of different sources till July 2016, including all the previous detections in the precedent phases for a total of 526.356 QSOs detected over 9376 deg$^2$ \citep{paris18}. These sources have been selected combining many color criteria at different redshifts from several surveys; in particular, to determine the redshift and the type of a source, the automated procedure of the SDSS makes a comparison between the observed spectra and a set of synthetic ones. At each supposed redshift, a least-squares fit to each
observed spectrum is done, using a general set of models for galaxies, QSOs, and cataclysmic variables. For each extragalactic model, a range of redshifts is explored and a chi-squared analysis is computed; the best five templates of each category are then stored in order of increasing reduced chi-squared. Finally, these latter models are re-fitted enhancing the resolution on the spectra (diminishing the initial velocity step); in this way, obtained the overall best fit, the class and redshift of each source are specified (for a more accurate description of the targets spectral classification and redshift measurement made by SDSS, see \cite{bolton12}). The spectroscopic data of the QSOs have been taken in the wavelength range $3600 \lesssim \lambda_{obs} \lesssim 10000$ {\AA}, with spectral resolution varying from $\sim 1300$ at 3600 {\AA} up to $\sim 2500$ at 10000 {\AA}. 

We have selected a sample of bright QSOs with $17.0 \leq I \leq 20.0$ ($-29.0 \lesssim M_{1450} \lesssim -26.0$) and $3.6 \leq z \leq 4.6$ from DR14 for a total of 2840 objects, analyzing their UV/optical spectra in order to characterize their physical properties and the surrounding IGM. The relative wide ranges in magnitude and redshift allowed us to investigate a possible luminosity and redshift evolution of our targets. In particular, the choice of this redshift coverage is mainly due to the following reasons: 1-QSOs at redshift $3.0 < z < 3.6$ present a bias against having $(u-g)<1.5$ for which sources with strong Lyman limit absorptions are mostly selected with respect to the other ones, altering the estimate of their physical properties in this redshift range \citep{prochaska09}. 2-At this epoch of the Universe, the transmission of the IGM is still relatively high, allowing the observations of the QSO characteristic features.


\section{The Method}

\subsection{Cleaning of the sample}

In order to proceed for a statistical analysis on the characterizing properties of our QSO sample, each spectrum has been individually explored, and several fundamental information extracted (i.e., the spectral slope, the free path of UV ionizing photons, the Lyman continuum escape fraction).

At first, a visual inspection of all the $3.6 \leq z \leq 4.6$ QSOs has been undertaken, so to obtain a least biased sample as possible; in this way, we have checked for the type of source and its redshift, automatically assigned by the SDSS pipeline. In fact, several targets have been wrongly classified as QSOs by the SDSS procedure, although they do not present any typical feature of a QSO UV/optical spectrum at these redshifts; for this reason, these spurious sources have been discarded from the final sample. Furthermore, a precise determination of the systemic redshifts of the sources is of
great importance in order to reduce the uncertainties on the estimate of the parameters describing the QSOs; therefore, each spectrum has been checked for inaccurate position of the strongest lines in emission of the QSO. To eventually refine the spectroscopic redshift
of a source, the OI emission line (1305.5 {\AA}, rest-frame) has been used; this line forms
in the outermost portion of the Broad Line Region (BLR) and is thus less affected by changing in its
spectral shape due to possible QSO outflows \citep{matsuoka05}. Taking the wavelength relative to
the OI line centroid ($\lambda_\mathrm{OI}$, in Angstrom), the new redshift has been obtained as $z_\mathrm{new} = \lambda_\mathrm{OI}/1305.5 -1$. Moreover, the modified redshifts of a few objects turned out to fall outside of the redshift interval used to select our QSOs. Excluding these latter sources and the spurious ones, the final sample consists of 2508 QSOs, each of which has been analyzed in order to obtain information about their evolution and the environment in which they reside.

\subsection{Estimating the spectral slope}\label{sub:spectral_slope}

The spectral slope of each QSO has been obtained by fitting the region red-ward of the Ly$\alpha$ line ($\sim$1216 {\AA} rest-frame) with the power law $F_\lambda = F_0 \lambda^{-\alpha_\lambda}$, and then extrapolating it in the bluer side of the spectra, imposing a softening at shorter wavelengths with $\alpha^{p}_\lambda=\alpha_\lambda-0.72$ at $\lambda_{rest}\lesssim 1000$ {\AA} \citep{cristiani16,stevans14,shull12,telfer02}.

\begin{figure}
\centering
\includegraphics[width=\columnwidth,angle=0]{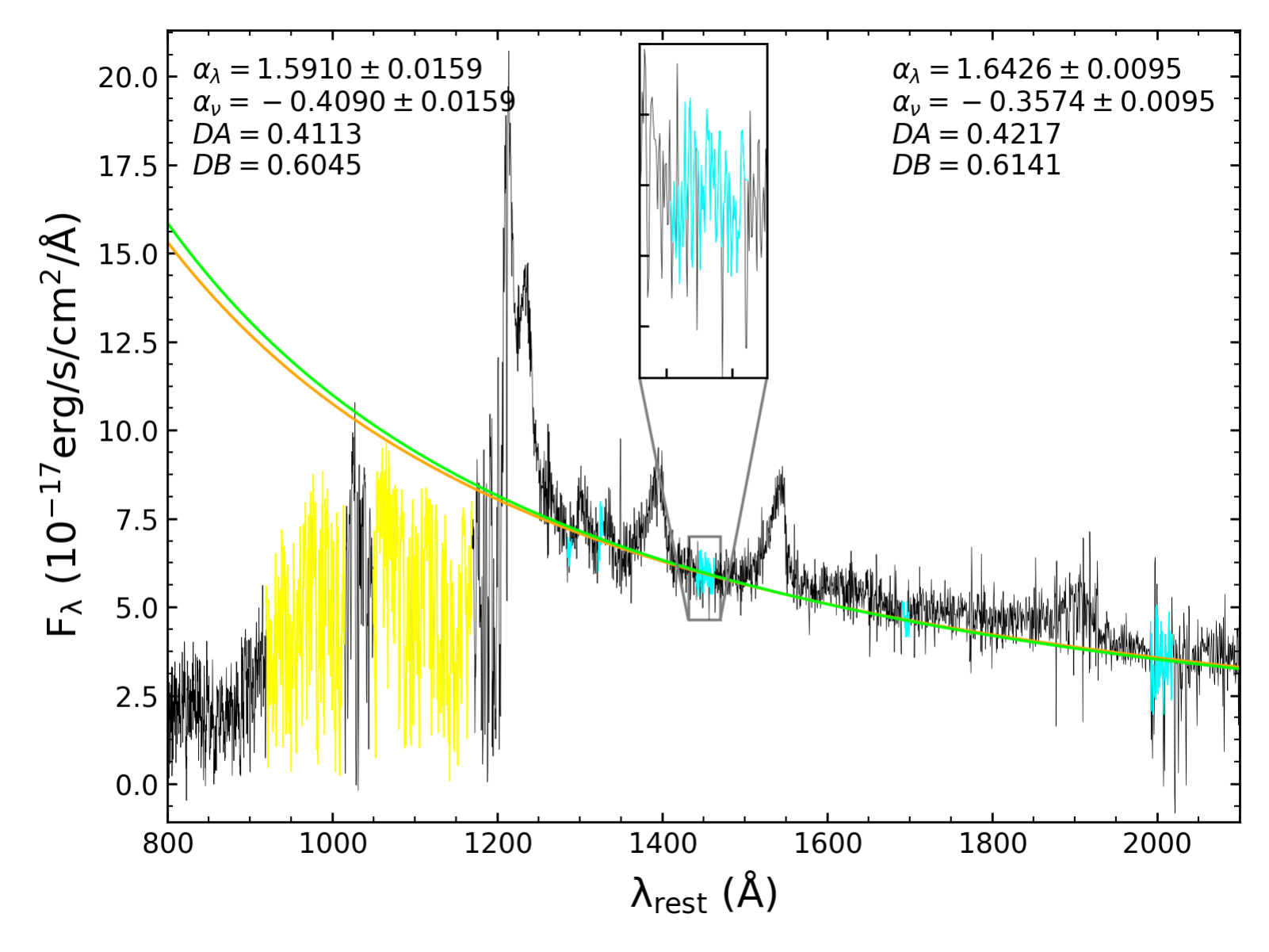}
\caption{Spectral slope and flux decrements for the QSO SDSS J105340.8+010335.7
at $z=3.66983$. The cyan portions of the spectrum are the five windows free of emission
lines. The yellow regions are the places of the spectrum in which the flux decrements
DA and DB have been
calculated. The solid orange and green lines represent the fit with and without the first
window (1284 - 1291 {\AA} rest-frame), respectively.}
\label{slope_example}
\end{figure}

At first, we have corrected each spectrum for interstellar extinction using the \cite{cardelli89} extinction law and the absorptions available in the SDSS database. Then, following \cite{cristiani16}, the five spectral windows free of emission lines, reported in their Table 1, have been used for the fit. In particular, to estimate the intrinsic continuum, the average values (both in flux and wavelength) obtained in each window, after an iterative 2$\sigma$ clipping, have been exploited. Leaving the normalization constant $F_0$ and the spectral slope $\alpha_\lambda$ as free parameters, a non-linear least-squares minimization routine has been used in order to obtain the best-fit parameters which described the local continuum of each QSO. Extrapolating the power law in the region blue-ward of the Ly$\alpha$ line, the flux decrements DA and DB \citep{oke1982} have also been obtained, in the spectral windows between the Ly$\alpha$ and Ly$\beta$ lines (1170 - 1050 {\AA}, rest-frame) and between the Ly$\beta$ line and the Lyman limit (1015 - 920 {\AA}, rest-frame), respectively. We have derived the flux decrements DA and DB in order to check the
possible dependencies of the escape fraction and of the free ionizing path from these (and other) physical parameters.
As an example, Fig. \ref{slope_example} shows the results of this procedure applied on the QSO SDSS J105340.8+010335.7
at $z_\mathrm{spec} = 3.66983$. Here, the cyan regions are the five spectral windows used to fit the continuum, where the third window (1440 - 1465 {\AA}, rest-frame) is zoomed to show the effect of the iterative 2$\sigma$ clipping. Moreover, in several cases, the first window (1284 - 1291 {\AA}, rest-frame) could be affected by the emission of the NV line, therefore the fit with and without this region has been computed (orange and green solid curves, respectively). The longest wavelength region, instead, could be affected by poor sky subtraction problems,
especially at high-z.
However, by default, the spectral slope for each QSO has been estimated considering the fit over all the spectral windows, in order to rely on all the exploitable data; in the upper left (right) side of Fig. \ref{slope_example} the values obtained for the all windows (no first window) case are shown. Finally, the yellow portions of the spectrum represent the regions in which the flux decrements DA and DB have been calculated.

Absorption features and/or noise in the spectra could even alter the slope of several sources, leading to an incorrect estimate of the intrinsic continuum of many QSOs. To avoid this kind of issue, the sources have been inspected one by one, removing the spectral windows affected by strong noise and recomputing the fit to estimate the spectral slope.  

\subsection{Estimating the Lyman Continuum escape fraction of SDSS QSOs at $z\sim 4$}\label{subsec:escape_frac}

The Lyman Continuum escape fraction for each QSO of our sample has been derived
following \cite{grazian18}. In summary, we estimate the mean flux
density short-ward, $F_\nu(900)$, and long-ward, $F_\nu(930)$, of the
Lyman limit (912 {\AA} rest-frame) and measure the escape fraction as
$f_{esc}=F_\nu(900)/F_\nu(930)$.

More precisely, the $F_\nu(900)$ is the mean flux density of the QSO
in the HI ionizing continuum region, between 892 and 905 {\AA} rest-frame, while $F_\nu(930)$ is the average flux density in the
non-ionizing region red-ward of the Lyman limit, between 915 and 960
{\AA} rest-frame. It is worth noting that this latter spectral window
has been slightly enlarged with respect to the one used by \cite{grazian18},
i.e., 915-945 {\AA}. In fact, after several tests, we have checked that with this new choice, the measurement of the mean flux density $F_\nu(930)$ for
our sources has turned out to be more stable than the previous one, being
less affected by possible absorption lines in the Lyman-$\alpha$ forest.
However, on average
the old and new escape fractions are quite similar to each other.
Finally, in calculating $F_\nu(930)$, the region between 935
and 940 {\AA} has been avoided, due to the Lyman-$\epsilon$ emission
line of the QSO. At $z\sim 3.6-4.6$, the adopted rest-frame
wavelengths 892-960 {\AA} correspond to observed wavelengths of
$\lambda_{obs}\sim 4100-5400$ {\AA}, which are well sampled by the
SDSS spectra we used in the present paper. In order to avoid the
presence of intervening absorbers in these regions, both $F_\nu(900)$
and $F_\nu(930)$ have been derived with an iterative 2$\sigma$ clipping
procedure.

It is worth noting here that we do not attempt to reconstruct the intrinsic QSO continuum, but we measure the observed mean flux level in these two spectral windows, assuming an equal residual Lyman series absorption above and below the Lyman limit. The observed distribution of the escape fraction could be
different from the intrinsic one due to the errors in the estimate
of the local continua, both red-ward and blue-ward of the Lyman
continuum. At this aim, we have carried out an estimate of the
flux variation in the Lyman forest. Precisely, for each QSO studied in
this paper, we measure the mean flux (after the 2$\sigma$ clipping
procedure) in the rest-frame wavelength 915-960 {\AA} and 960-1005 {\AA} and
compute their flux ratio. It turns out that this flux ratio does not
depend on the absolute magnitudes or spectroscopic redshifts, and its
scatter is $\sim$0.15 dex. We assume that the same error can be applied
when we estimate the individual QSO continua (both red-ward and blue-ward
of the Lyman continuum). Just for comparison, \citet{prochaska10}
estimated a 10\% uncertainty on the estimate of the SDSS QSO continuum in the Lyman forest.
Adopting this $\sim$0.15 dex continuum uncertainty, we find that the
typical scatter of the escape fraction is $\sim$15-17\% for $f_{esc}>75\%$, while it is
progressively decreasing to 1-2\% at $f_{esc}<20\%$. Moreover, the output
values of the escape fraction are fully consistent with the input values, without
any systematic. We have checked that the mean value of the escape
fraction distribution is not perturbed significantly by the scatter
due to flux ratio uncertainties. Interestingly, the intrinsic
distribution (i.e., after deconvolution by the observed scatter) gives
slightly higher mean values of $f_{esc}$ than the observed distribution. In
the following, given that the (reconstructed) intrinsic distribution is
too noisy, we prefer to plot only the observed distribution.

As discussed in \cite{grazian18}, this technique is equivalent to measure $f_{esc}=\mathrm{exp}(-\tau_\mathrm{LL})$, where $\tau_\mathrm{LL}$ is the average opacity at the Lyman limit. It could include the cumulative effect of
small absorbers ($\tau_\mathrm{LL}<2$), which can not be individually detected in the SDSS QSO spectra \citep{prochaska10}.

We avoid using the spectral window close to the 912 {\AA}
rest-frame of the QSO (i.e., 905-912 {\AA} rest-frame) since this
region can be affected by the QSO proximity effect. Moreover, the
uncertainties in spectroscopic redshift determination of SDSS spectra,
typically of $\sigma_z\sim 0.01$, does not allow us to locate
precisely the position of the Lyman limit, so we decide to place our spectral
window far from this region.

It is worth noting here that our estimate of the LyC escape fraction
$f_{esc}$ for the SDSS QSOs at $z\sim 4$ can be considered a lower limit to
the real value for several reasons: 1-we do not correct this value for
the spectral slope of the QSOs. In the $\lambda_{rest}\le 1200$ {\AA}
region, a typical value for the spectral slope is $\alpha_\nu\sim
-1.7$ \citep{lusso15}. Correcting for this slope, the true values
of $f_{esc}$ for our QSOs should be revised upward by $\sim 6\%$.
2-There are possible Lyman-$\alpha$ absorbers which serendipitously
fall between the two spectral windows adopted to estimate $f_{esc}$,
i.e., 905-915 {\AA} in the QSO frame.
Following \citet{inoue14}, we compute a flux decrement of $\sim 0.6$ between 930 and 900 {\AA} rest-frame, due to
intervening IGM (see their Fig. 4 for $z_{spec}=4.0$).
3-We exclude from
the escape fraction calculations the proximity region of the QSO,
which is instead a clear signature of the ionization of the surrounding IGM. 4-As we will discuss in Section \ref{subsec:LyCvsMFP}, if we estimate the mean flux density $F_\nu(900)$
in a shorter spectral window between 898 and 905 {\AA} rest-frame, the escape fraction turns out to be slightly larger than the one computed in the wider region between 892 and 905 {\AA} rest-frame.

\subsection{Estimating the Individual Free Path of HI ionizing photons}\label{subsec:mfp_estimate}

The free path of photons at 912 {\AA} rest-frame is a fundamental parameter
in order to estimate the HI photo-ionization rate. The IFP is defined as
the physical distance a 912 {\AA} rest-frame photon can travel before being attenuated by a factor of
$e^{-1}$ in flux. Such an attenuation can be caused by
encountering a Lyman limit
system with optical depth $\tau_\mathrm{HI}=1$ or, alternatively, by the cumulative effect of a great number of more transparent absorbers, with $\tau_\mathrm{HI}<<1$, even if it is not possible to detect them individually.

\begin{figure}
\centering
\includegraphics[width=\columnwidth,angle=0]{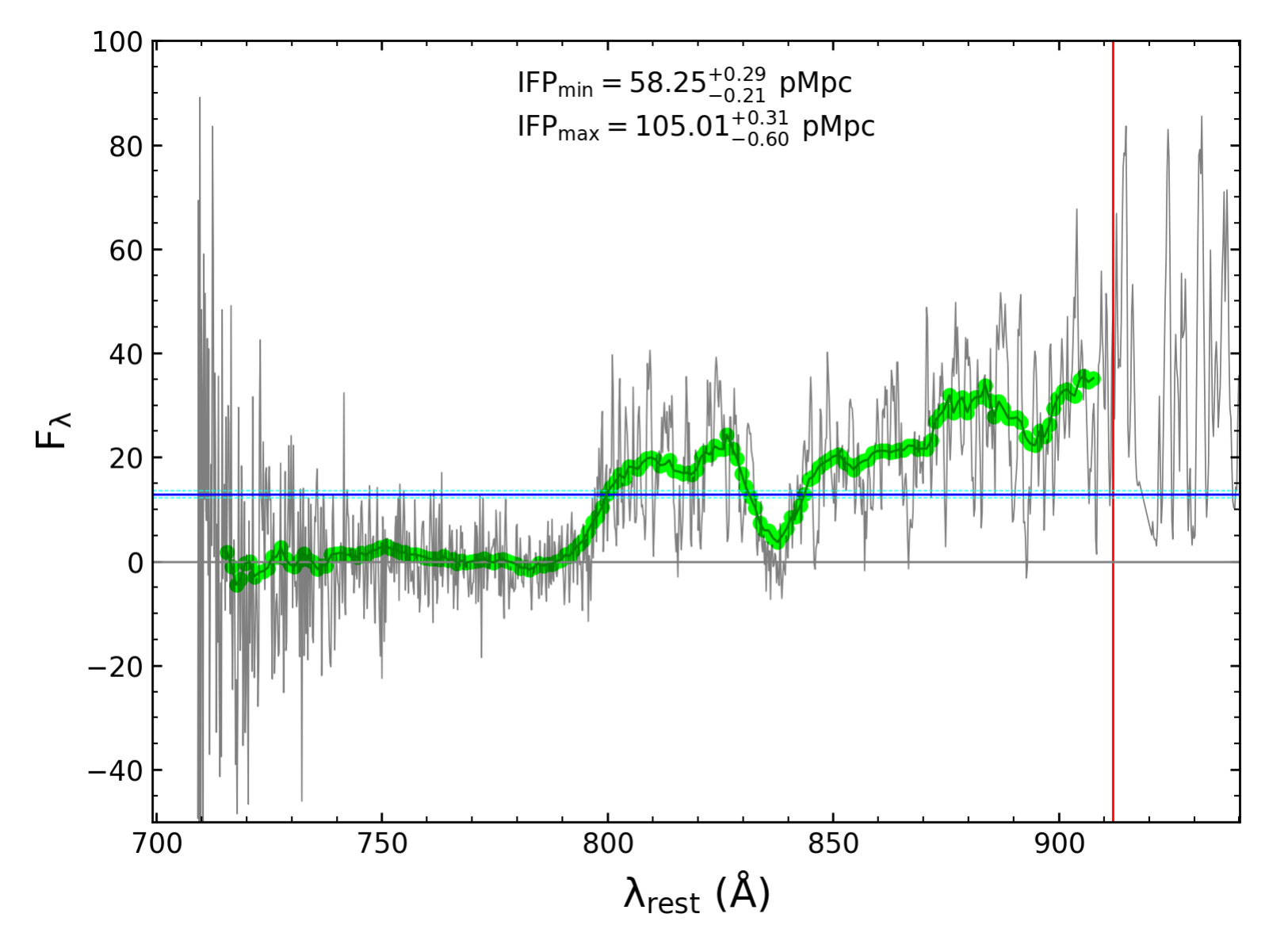}
\caption{
Individual free path estimate for the QSO SDSS J114514.2+394715.9
at $z_\mathrm{QSO} = 4.06557$. The green circles and solid line represent the average fluxes in the 10 {\AA} windows and the spline, respectively. The red line marks the Lyman limit. The horizontal blue and cyan lines represent $f_{red}$ and its uncertainty, respectively. In this case, the choice of IFP corresponds to $\lambda_{rest}\sim 800$ {\AA}. The flux on the y axis is in arbitrary unit.}
\label{fig:MFP4654}
\end{figure}

In order to carry out a statistical analysis, the IFP has been estimated individually for each spectrum. At first, the SED of each source was corrected for its intrinsic spectral slope, considering a break in the continuum at $\sim$ 1000 {\AA} rest-frame, as found by \citet{stevans14,shull12}. This has been done flattening the spectral slope of each QSO (obtained as described in Section \ref{sub:spectral_slope}) for the value $\Delta \alpha_\nu = 0.72$ \citep{cristiani16} at $\lambda\le$ 1000 {\AA} rest-frame. Starting from $\lambda_{in} = 3600/(1+z)$, the region blue-ward of the Lyman limit has been divided into several spectral windows of width $\Delta \lambda$ and upper limit $\lambda_{fin} = \lambda_{in} + \Delta \lambda$. By default, $\Delta \lambda$ has been chosen to be 10 {\AA}, being this value a good compromise in following the spectral features of the specific QSO without under/overestimating the IFP; in this way, the spectral window has been moved upward of 1 {\AA} in every cycle until the Lyman limit, i.e., $\lambda_{fin} = 912$ {\AA}, has been reached. In each window, the average fluxes and wavelengths have been calculated through
an iterative $2\sigma$ clipping on the respective data, and then interpolated with a spline. In this case, an asymmetric clipping has been computed in order not to underestimate the continuum level of the spectrum: all the negative residuals have been rejected. The average flux obtained in the upper window near the Lyman limit, $f_{912}$, has been used to compute the flux reduced by a factor $e^{-1}$, as $f_{red} = f_{912}/e$. The intersection between the spline and $f_{red}$ defines the wavelength relative to the (Lyman limit) absorber along the line of sight, $\lambda_\mathrm{LLS}$, whose redshift has been obtained as $z_\mathrm{LLS} = \lambda_\mathrm{LLS}/912-1$.
Therefore, the comoving distances between the absorber and the source have been computed as $D_c = D_c(z_\mathrm{QSO}) - D_c(z_\mathrm{LLS})$, where $z_\mathrm{QSO}$ is the redshift of the QSO. Finally, the free path of ionizing photons is the proper distance between the QSO and the LLS, and it has been derived as IFP $=D_p^\mathrm{LLS}=D_c/(1+z_\mathrm{LLS})$.
We have checked that the IFP value does not depend on the choice of
$\Delta \alpha_\nu = 0.72$. We have derived the individual IFPs for all the QSOs with $\Delta \alpha_\nu = 0.5$ and $\Delta \alpha_\nu = 1.0$, which are still allowed by the uncertainties provided e.g., by \citet{stevans14,shull12,telfer02} and found that the effect is negligible, with a
difference of $\sim 0.1$ pMpc on the mean value of IFPs.

In several cases, absorption features blue-ward of the Lyman limit could drag the interpolated fluxes below $f_{red}$, and then bring it above this threshold; this trend could create many intersections of the spline with $f_{red}$ for each of which the individual free path has been computed. Both the minimum and maximum IFP values have been derived, as can be seen in Fig. \ref{fig:MFP4654} for the QSO SDSS J114514.2+394715.9
at $z_\mathrm{QSO} = 4.06557$. For all our QSOs, the real individual free path of ionizing photons is clearly associated to the maximum IFP value, being the minimum one just the result of an absorption feature; for this reason, the maximum IFP has been set as the definitive value for each QSO by default.


\section{Results}

\subsection{Spectral slope distribution}

To study the ensemble properties of our QSO sample, we have at first obtained the spectral slope distribution in order to carry out a statistical analysis of this parameter and to explore a possible evolution in magnitude and/or redshift.

\begin{figure}
\centering
\includegraphics[width=\columnwidth,angle=0]{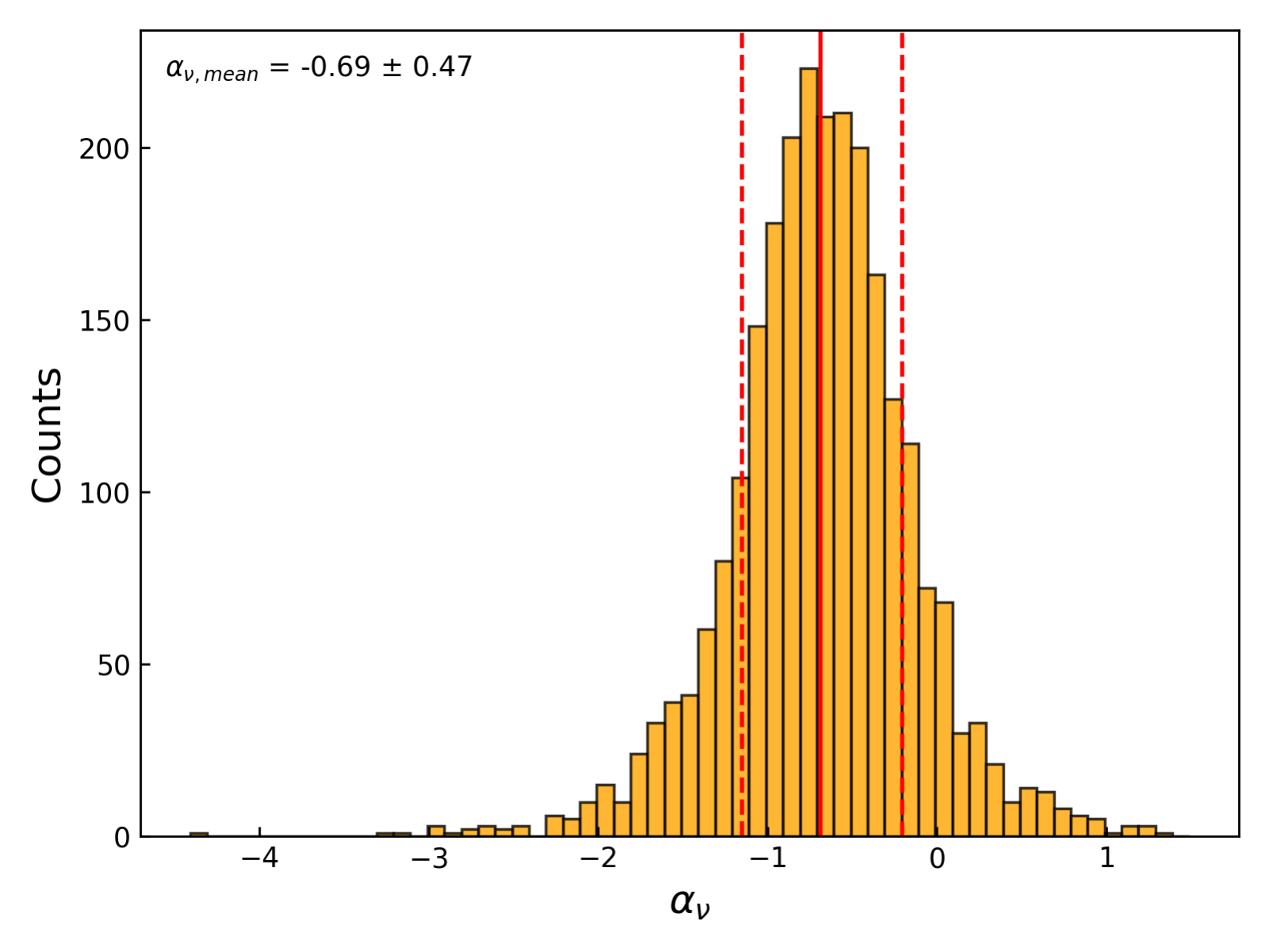}
\caption{Spectral slope distribution. The vertical red solid and dashed lines are the mean value and its uncertainties, respectively.}
\label{fig:slope_distr}
\end{figure}

Fig. \ref{fig:slope_distr} shows the distribution function with the red line marking the spectral slope mean value, $\alpha_{\nu}=-0.69 \pm 0.47$ (i.e., $\alpha_\lambda \approx 1.31$), and where the uncertainties (red dashed lines) have been computed as half of the difference between the 84th and 16th percentiles. This result is in agreement with previous estimates of \cite{cristiani16}, who found $\alpha_\lambda = 1.30$ at $3.6 < z < 4.0$, and with those of \cite{telfer02} and \cite{stevans14} who obtained $\alpha_\lambda \approx 1.31$ and 1.17 at lower redshift, respectively.

To investigate the evolution of the spectral slope, the whole QSO sample has been split in two halves including fainter and brighter sources with respect to different absolute magnitude thresholds $M_{1450}$ (from $\approx -29.0$ up to $\approx -26.0$); the two distributions corresponding to each $M_{1450}$ have been compared using a two-sample Kolmogorov-Smirnov (KS) test, choosing to reject the null hypothesis (i.e., the hypothesis that two independent samples are drawn from the same parent distribution) for a p$_\mathrm{value}$ below 1 $\%$. With this test, a luminosity evolution of the spectral slope has been rejected.

\begin{figure}
\centering
\includegraphics[width=\columnwidth,angle=0]{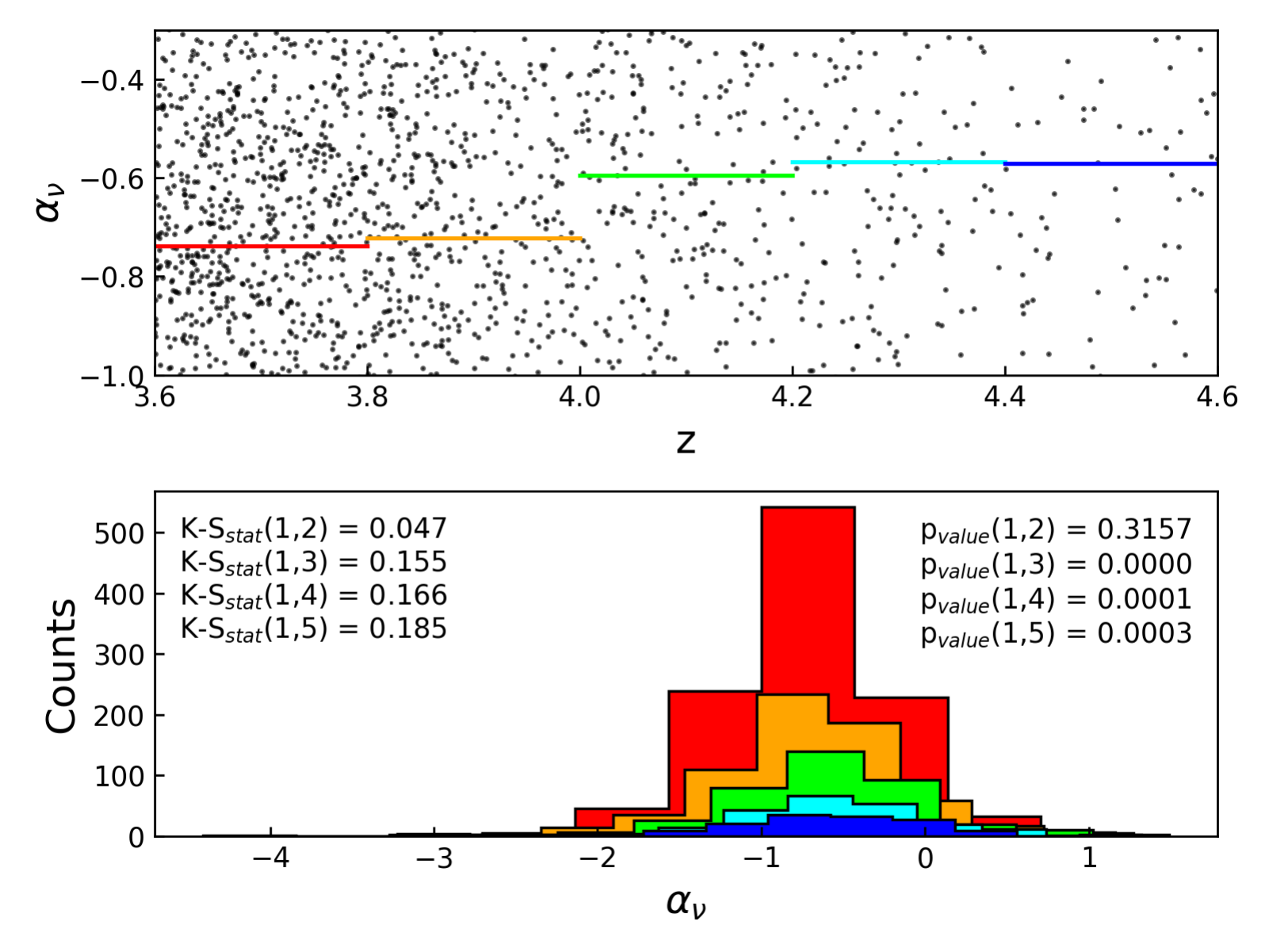}
\caption{{\em Upper panel}: spectral slope as a function of redshift. The horizontal red, orange, green, cyan, and blue lines are the average values from the lowest to the highest redshift bins. {\em Bottom panel}: distribution of the spectral slope in each redshift bin (with the same color of the upper panel).} 
\label{fig:slope_z}
\end{figure}

The same procedure has been used to study the evolution of $\alpha_\nu$ with redshift. In this case, the QSO sample has been split in five redshift bins having $\Delta z = 0.2$, in order to test possible differences among the various distribution functions, as shown in Fig. \ref{fig:slope_z}. The horizontal solid lines in the upper panel of the Figure represent the average values of the spectral slope in each bin of redshift, increasing from $\alpha_\nu \approx -0.74$ in the first bin up to $\alpha_\nu \approx -0.57$ in the last one. The corresponding distributions are plotted in the lower panel where the KS tests between the first and the other redshift bins are also shown. It is interesting to note that the KS tests between the first and the three last distributions are significant, defining a possible redshift evolution of the analyzed parameter.

This is probably caused by the shorter range in wavelengths used to compute the power-law fit which affects the QSO spectra at high redshift, i.e., the fact that the reddest spectral window in the spectra could exit the wavelength coverage of the SDSS for these sources. In order to test this hypothesis, we have recomputed the spectral slopes for our QSO sample excluding the spectral window at the highest wavelengths when fitting the power law. In this case we obtain an increase of $\alpha_\nu$ from $\approx -0.65$ at $z \sim 3.7$ to $\approx -0.57$ at $z \sim 4.5$. Thus, at low redshift ($z \lesssim 3.9$), the average value of the slope is slightly higher than the one obtained with all the free emission windows, while it remains unchanged at high redshift. In this case, the KS test is no more significant within the different bins in redshift. For these reasons, the above-mentioned redshift trend of the spectral slope seems to be mostly caused by a systematic effect due to the reduced spectral range, rather than to a physical evolution.

\subsection{Lyman Continuum escape fraction distribution}

\begin{figure}
\centering
\includegraphics[width=\columnwidth,angle=0]{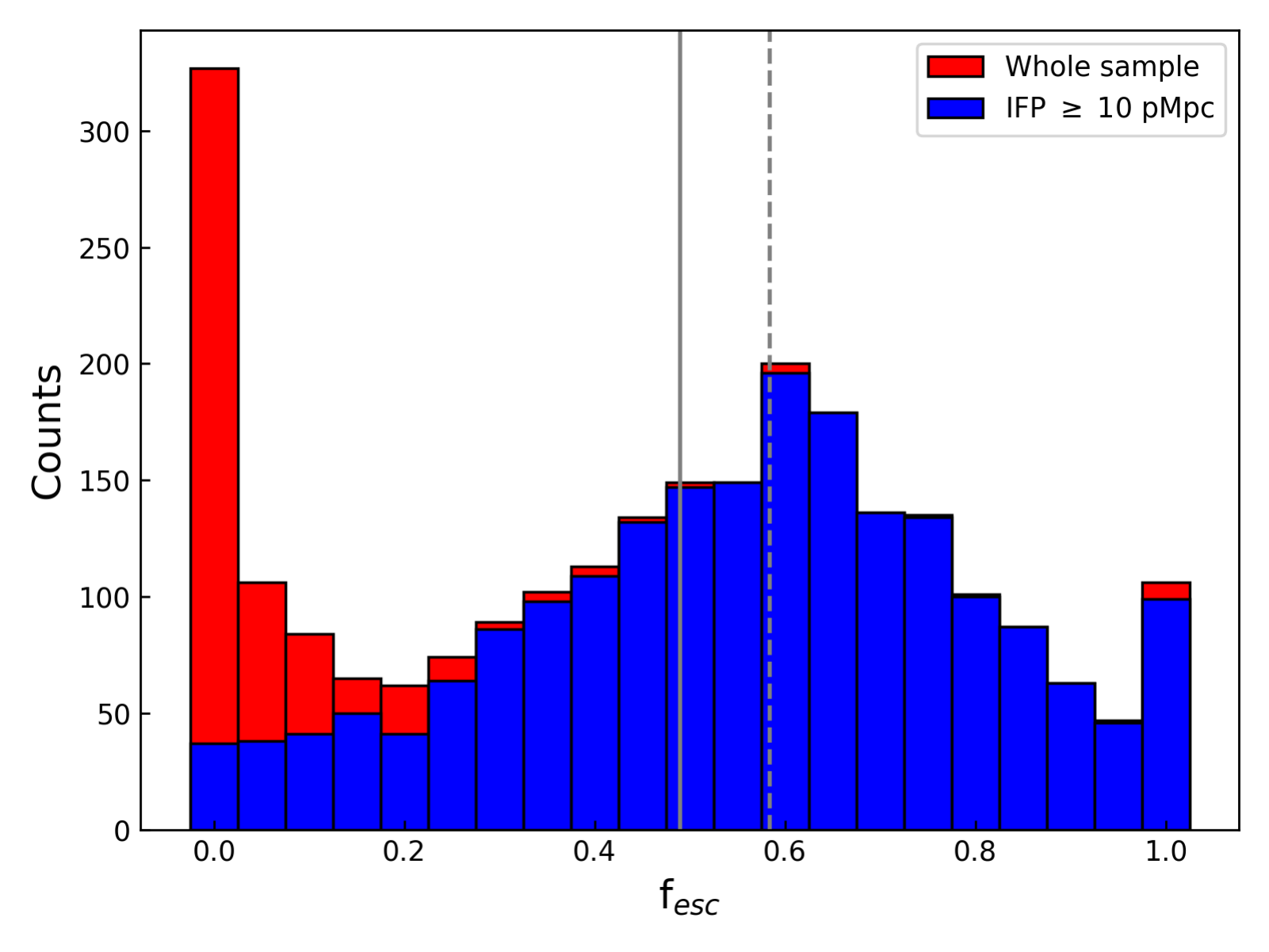}
\caption{Lyman Continuum escape fraction distribution for the whole sample (in red) and for sources with IFP $\geq$ 10 pMpc (in blue). The solid and dashed gray lines represent the mean values of the former and latter distributions, respectively.}
\label{fig:fesc}
\end{figure}

Depending on the intrinsic properties of each specific QSO
and on its possible absorption features, the computed LyC escape fractions cover a broad range of values,
going from approximately zero to 100$\%$. We have found an average value $f_{esc} = 0.49 \pm 0.36$ in our sample, as reported in Fig. \ref{fig:fesc} which shows the escape fraction distribution function of all our sources (red histogram); this one results in a multi-modal distribution with a peak at low values (14.5\% of the QSOs with $f_{esc}<0.10$), a minimum between $0.10\le f_{esc}\le 0.3$ and a broad dispersion at higher values, as also found by \cite{cristiani16}. Also shown in this Figure is the distribution from the sub-sample excluding sources with IFP < 10 pMpc, resulting in an average value of $f_{esc} = 0.58 \pm 0.25$ (in blue, see Sub-section \ref{subsec:LyCvsMFP}). In particular, individually inspecting the sources within the first bin, the low $f_{esc}$ peak has been attributed to the presence of several QSOs with LLSs
which are intrinsic to the QSO itself or intervening along the line of sight, which absorb almost all the UV radiation escaping from them.
This causes a rapid decline of the emitted flux in the region blue-ward of the Lyman limit producing a negligible escape fraction (see Sub-section \ref{subsec:z_MFP_evol} for a discussion on the effect of these sources on the final statistical analysis of the QSO physical parameters).

As also done for the spectral slope, the dependence of the LyC escape fraction on both redshift and magnitude has been tested by splitting the whole sample in sources with lower and higher redshifts ($z \leq 4.1$ and $z > 4.1$), and QSOs brighter and fainter than a given absolute magnitude threshold, respectively. The different populations of objects have been compared with a KS test whose results have allowed us to exclude a possible evolution of $f_{esc}$ with redshift. Regarding the
dependency of the escape fraction on the magnitude, splitting the sample in different magnitude bins, we have obtained $p_\mathrm{value} = 0.007$ in the brightest bin ($M_{1450}$ < -27.4) which could be significant for a possible trend of the escape fraction with luminosity. However, at fainter magnitudes, $p_\mathrm{value}$ increases to values higher than 1$\%$, not allowing to confirm an evolution of $f_{esc}$ with luminosity.

\subsection{Individual Free Path distribution}

\begin{figure*}
\centering
\includegraphics[width=16cm,angle=0]{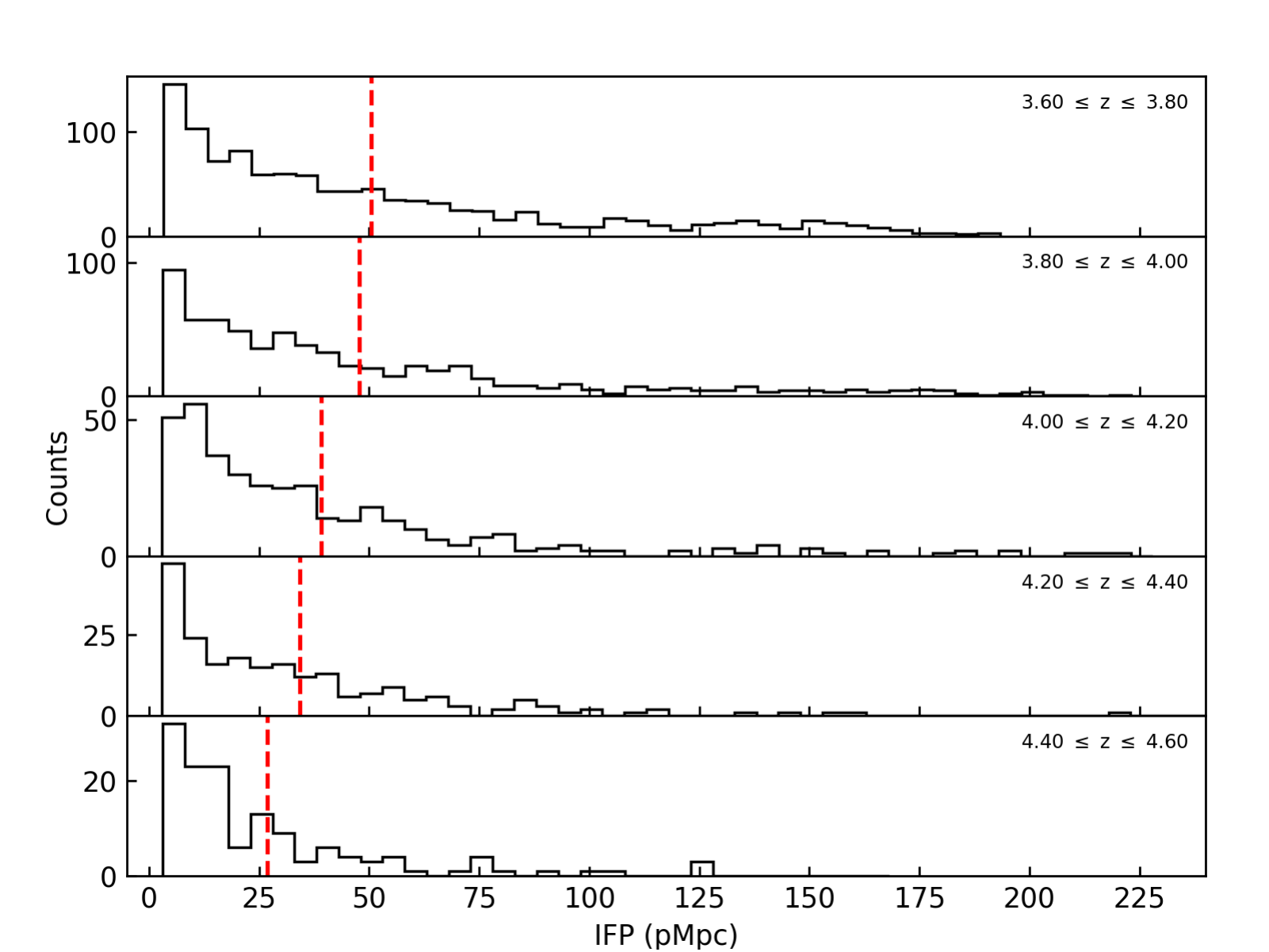}
\caption{Free path distribution functions for the whole QSO sample at $3.6\le z\le 4.6$ in redshift bins of $\Delta z = 0.2$. The dashed red lines represent the mean value in each
redshift bin.}
\label{fig:MFP_distr}
\end{figure*}

The analysis implemented in this paper has led to estimate, for the first time, the mean free path of UV ionizing photons of a QSO sample at redshift $z \sim 4$ in a statistical way, i.e., from the distribution functions of IFPs. These ones are displayed in Fig. \ref{fig:MFP_distr} for redshift bins of $\Delta z = 0.2$, where the dashed red lines mark the mean value in each bin; a peak in the distributions is present at low free paths, with a broad tail at higher values, and a mean free path MFP $\sim 43$ pMpc at z $\sim$ 4.

It is well known that this parameter evolves in redshift, being the Universe increasingly more and more composed of neutral hydrogen which absorbs UV photons approaching the EoR (see Sec. \ref{subsec:z_MFP_evol}).
At $z<5$ the HI-ionizing background
is quasi-uniform (e.g., Meiksin \& White 2004;
Becker et al. 2015), resulting in a skewed but uni-modal
distribution of free paths that reflects Poisson variance in the incidence of
(partial) Lyman limit systems along individual lines of sight.
The dependency of the free paths with the luminosity of the QSOs will be discussed in the next section.


\section{The Free Path versus LyC Escape Fraction Connection}\label{sec:MFP_vs_esc}

\subsection{Dependence of LyC escape fraction on QSO Individual Free Path}\label{subsec:LyCvsMFP}

\begin{figure}
\centering
\includegraphics[width=\columnwidth,angle=0]{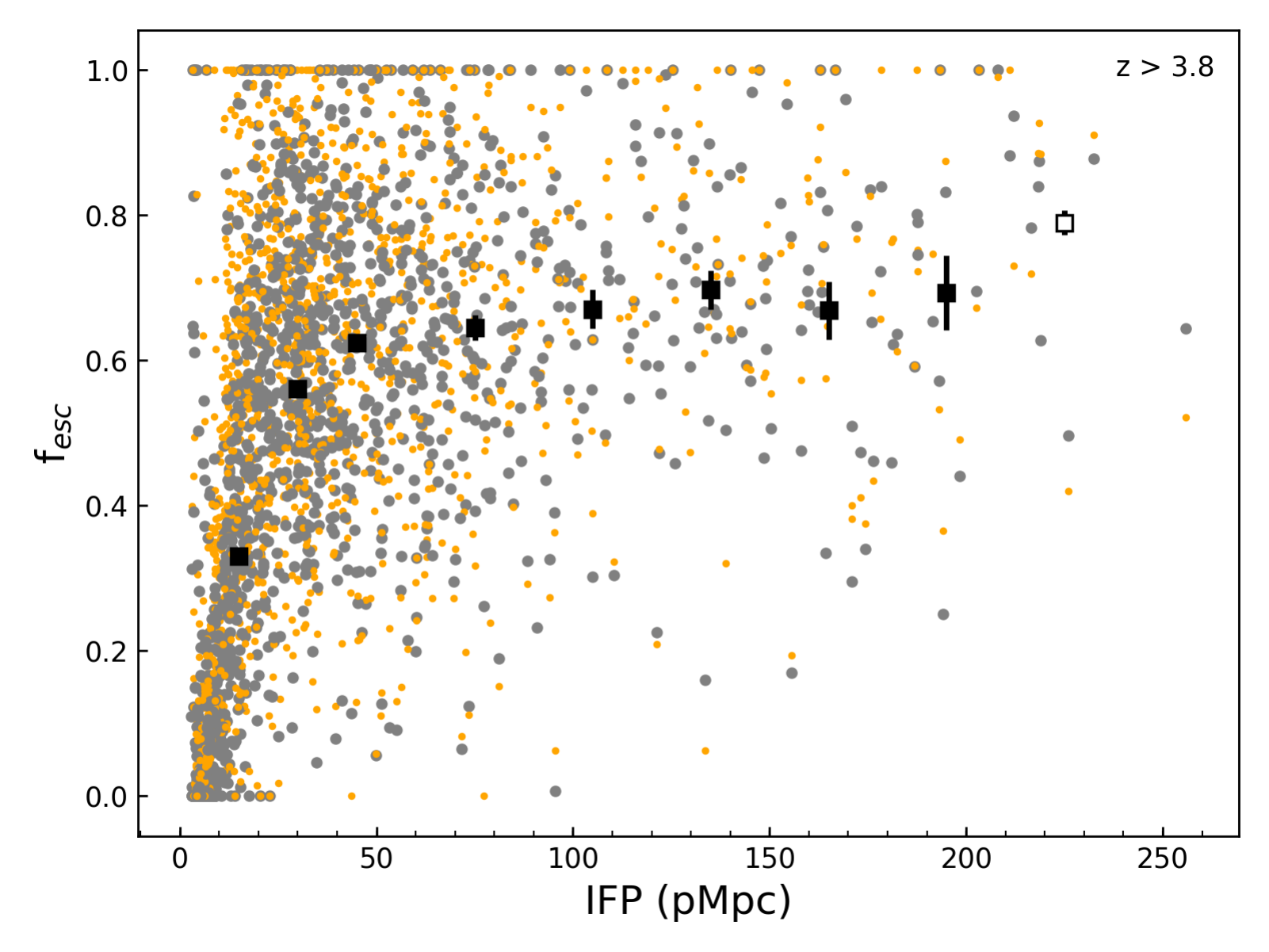}
\caption{Escape fraction vs individual free path for QSOs at z > 3.80. The gray points show the escape
fractions calculated with the ionizing window 892-905 {\AA} rest-frame; the orange points
represent the
escape fractions with the reduced ionizing window 898-905 {\AA} rest-frame. The black filled square represent the mean $f_{esc}$ in IFP bins of 30 pMpc (the open square is the mean of $f_{esc}$ in IFP bin containing less than 10 sources); the vertical bars are the error on the mean in each bin of IFP. QSOs with $z<3.8$ have been excluded here to avoid possible biases due to their IFP under-estimate, as discussed in Section \ref{subsub:limited_cov}.
}
\label{fescMFP}
\end{figure}

The distribution of the LyC escape fraction of our QSOs is multi-modal,
with a narrow peak at $f_{esc}\lesssim 0.2$ and a broad distribution from
0.3 to 1.0, as discussed in the previous section. In order to
investigate the origin of this multi-modality, we check for any dependencies
of $f_{esc}$ on the other physical parameters. No dependence has been found
comparing against spectral slope $\alpha_\nu$, redshift, absolute
magnitude $M_{1450}$, and Lyman decrements (DA, DB). The only
exception is against the individual free paths, as shown in
Fig. \ref{fescMFP}. It is interesting to note the trend of the
escape fraction with the individual free path, at IFP $\le$ 20 pMpc. In
particular, a deficiency of high escape fractions at low IFP and a
lack of low $f_{esc}$ at high free paths are evident. Moreover, an
increase of the escape fraction proportional to the individual free path
seems to be present up to IFP of the order of $\sim 10-20$ pMpc.

\begin{figure}
\centering
\includegraphics[width=\columnwidth,angle=0]{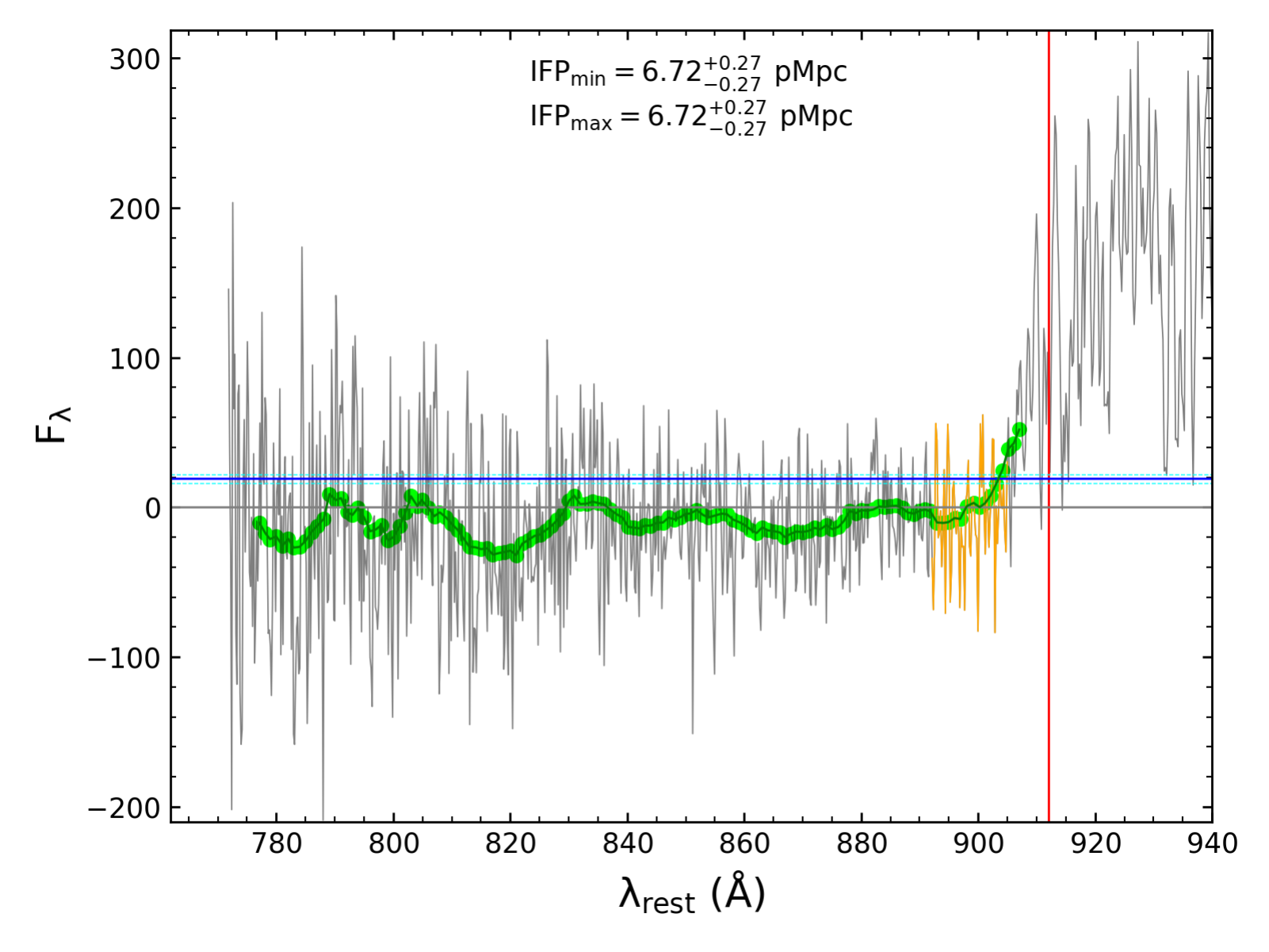}
\caption{
Free path estimate for the QSO SDSS J034402.9-065300.6
at $z=3.92455$,
a QSO with short IFP and low value of escape fraction, thus representing
an example of correlation between IFP and $f_{esc}$. The orange spectral region
marks the ionizing window used to calculate the escape fraction (i.e.,
892-905 {\AA} rest-frame).
}
\label{ex2}
\end{figure}

About the lack of sources at low free paths and high escape
fractions, this could be interpreted as an effect due to the choice of
the spectral window in the ionizing region, between 892 and 905 {\AA} rest-frame, adopted
in the $f_{esc}$ derivation. Figure \ref{ex2} shows the estimate of the
IFP for the QSO SDSS J034402.9-065300.6
at $z=3.92455$. This source has
IFP = 6.7 $\pm$ 0.3 pMpc and $f_{esc}=0.0$, thus consistent with the above
consideration. In this case, the wavelength associated with the individual free
path is inside the ionizing window of the escape fraction (highlighted
in orange in the Figure), creating a correlation between the two
parameters. At the same time, a reduced spectral window, closer to
the Lyman limit at 912 {\AA}, could increase the average flux in this
region, leading to an enhanced escape fraction for the specific
QSO. As an example, the rest-frame 892-905 {\AA} window was reduced
to 898-905 {\AA} and the escape fractions recomputed using this modified window for the whole sample.
This is displayed in Figure
\ref{fescMFP}; here, the escape fractions adopted in this paper (gray points, calculated as explained in Section \ref{subsec:escape_frac}) and the modified ones (orange points) are plotted for each source at $z>3.80$ (see Section \ref{subsub:limited_cov}) as a function of their free paths.
As can
be seen, the orange points tend to populate the low
IFP region at high escape fractions, reducing the previously discussed
deficit of these sources. This confirms that the lack of these QSOs is
mostly due to the spectral window (892-905 {\AA} rest-frame) adopted for calculating the escape fraction
rather than to a physical mechanism. In principle, adopting an even shorter
window close to 912 {\AA} rest-frame for the $f_{esc}$ derivation would further reduce the number of sources with low IFP and low $f_{esc}$ but, at the same time, the scatter would significantly increase. For these reasons, we decide to fix the spectral windows for the escape fraction measurements to the rest-frame 892-905 {\AA} wavelength range.
Interestingly, \citet{cristiani16} adopted a
wavelength interval much further than the one adopted here (850-880 {\AA}
rest-frame), and their peak at low $f_{esc}$ is slightly enhanced compared to our. At $f_{esc}\ge 0.20$ \citet{cristiani16}
have a fraction of 69.5\% of their sample, while we have 78.8\% of the QSOs above this threshold.

In Fig. \ref{fig:fesc} we plot also the escape fraction distribution (in blue) in case the sources with short IFP have been removed. Limiting the sample to IFP $\ge 10$ pMpc, the
peak at $f_{esc}\sim 0$ disappears, and a continuous trend
is in place, indicating that the sources with low escape fraction are probably due to absorbers falling by chance at small distance from the QSO, or associated to it.
In the following sections we discuss about the nature of these absorbers.

Another interesting feature emerges from Fig. \ref{fescMFP}: excluding very few objects with $f_{esc}\sim 20\%$ and large IFP, a trend of larger escape fraction appears for longer IFP, as highlighted by the mean of $f_{esc}$ in IFP bins of 30 pMpc (black filled and open square) and by the lower envelope of the individual QSOs (gray points). In particular, at very large IFP ($>170$ pMpc), only few QSOs have $f_{esc}\le 40\%$. We have checked that this correlation does not depend on the choice of the parameter $\Delta \alpha_\nu = 0.72$ adopted. The observed trend remains if we change this parameter to 0.5 or 1.0. This correlation, however, could be limited only to few QSOs with IFP greater than 170 pMpc, while for the bulk of the QSOs with $30<\mathrm{IFP}<150$ pMpc the escape fraction shows no strong dependency on the IFP, remaining almost constant at $\sim 60-70\%$. Finally, we have also computed a KS test to check whether the sample of QSOs with $\mathrm{IFP}>150$ pMpc and the one with $30<\mathrm{IFP}<150$ pMpc may be drawn by the same parent population. We find a p$_\mathrm{value}$ of 0.064, thus we can conclude that the differences between the two
samples are not significant. At present, however, we can not exclude that this trend is due to some systematic effects of the data (e.g., due to the color selection of $z\sim 4$ QSOs). More checks are needed in the future to substantiate such statement.

In the following sections, when dealing with the IGM properties, we will consider two extreme options to interpret the nature of the QSOs with IFP $\le 10$ pMpc: 1-all the sources with IFP $\le 10$ pMpc are affected by intervening absorbers (option 1). 2-all these QSOs are instead related to associated (intrinsic) absorption (option 2), due to the presence of dense neutral absorbers in their vicinity. In the former, the IFP is the physical distance between the emitter and the absorbing system, while in the latter the IFP reflects the outflow velocity of material which sits close to the QSO. In "option 1", the QSOs with IFP $\le 10$ pMpc have been included in the statistical analysis of the probability distribution function for the free paths. In "option 2", instead, they have been excluded not representing the mean properties of the IGM, since the central engines are affected by their surrounding Circum Galactic Medium (CGM) or host galaxy. There could also be associated absorbers with optical depth $\tau\le 1$, which would mildly suppress the flux blue-ward of the Lyman limit, thus mimicking intervening absorbers with IFP $\ge$ 10 pMpc. We assume here that all the systems with IFP $\ge$ 10 pMpc are intervening absorbers. This assumption will reduce the observed IFP w.r.t. the intrinsic one. 
In Section \ref{subsec:absorbers} we discuss in detail the nature of the close absorbers with IFP $\le 10$ pMpc.

\subsection{The redshift evolution of the Mean Free Path}\label{subsec:z_MFP_evol}

Since our data cover a relatively wide redshift range, we investigate
here the evolution of the mean free path of ionizing photons with the
cosmic age. An evolution in redshift of the mean free path is
certainly expected, due to the properties of the IGM and its evolution
with the cosmic expansion.
It is indeed known that this parameter should evolve in
redshift, being the Universe increasingly composed by neutral hydrogen
which absorbs UV photons approaching the EoR.
A precise estimate of the redshift evolution of MFP is fundamental, since it is one of the primary
sources of uncertainty in the calculation of the photo-ionization rate
$\Gamma_\mathrm{HI}$, as we will discuss in the next sub-section.

In the past, the mean free path of ionizing photons has been estimated
mainly by the number statistics of LLSs with
redshift (e.g., Songaila \& Cowie 2010; Prochaska et al. 2013;
Inoue \& Iwata 2008; Inoue et al. 2014) or by stacking
a large number of QSO spectra at the same redshifts (e.g., Prochaska
et al. 2009; Worseck et al. 2014). Given the high quality of the SDSS spectra
analyzed here for the escape fraction determination, we decided to
measure the free path for each individual QSO at $3.6 \leq z \leq 4.6$ in order to
derive the mean value and scatter of this parameter, leading to a statistical analysis of the free path distribution.
Moreover, the PDF of IFP gives better complementary information than a single average value (MFP).

At the beginning, we compute the individual free paths for SDSS QSOs at
$3.6 \leq z \leq 4.2$ and $I\le 19.5$ and find that the mean values of the
IFP distribution are larger than the MFP obtained by \citet{prochaska09} and
\citet{worseck14} through stacking technique. We then extend the
magnitude limit of our QSO sample to $I=20.0$ and derive mean values of
IFP which are consistent with the \citet{prochaska09} and
\citet{worseck14} results. We then check the dependence of the IFP
from the absolute magnitude $M_{1450}$ of the SDSS QSOs, since we have
indications of a possible trend of the IFP with QSO luminosity.

\begin{figure}
\centering
\includegraphics[width=\columnwidth,angle=0]{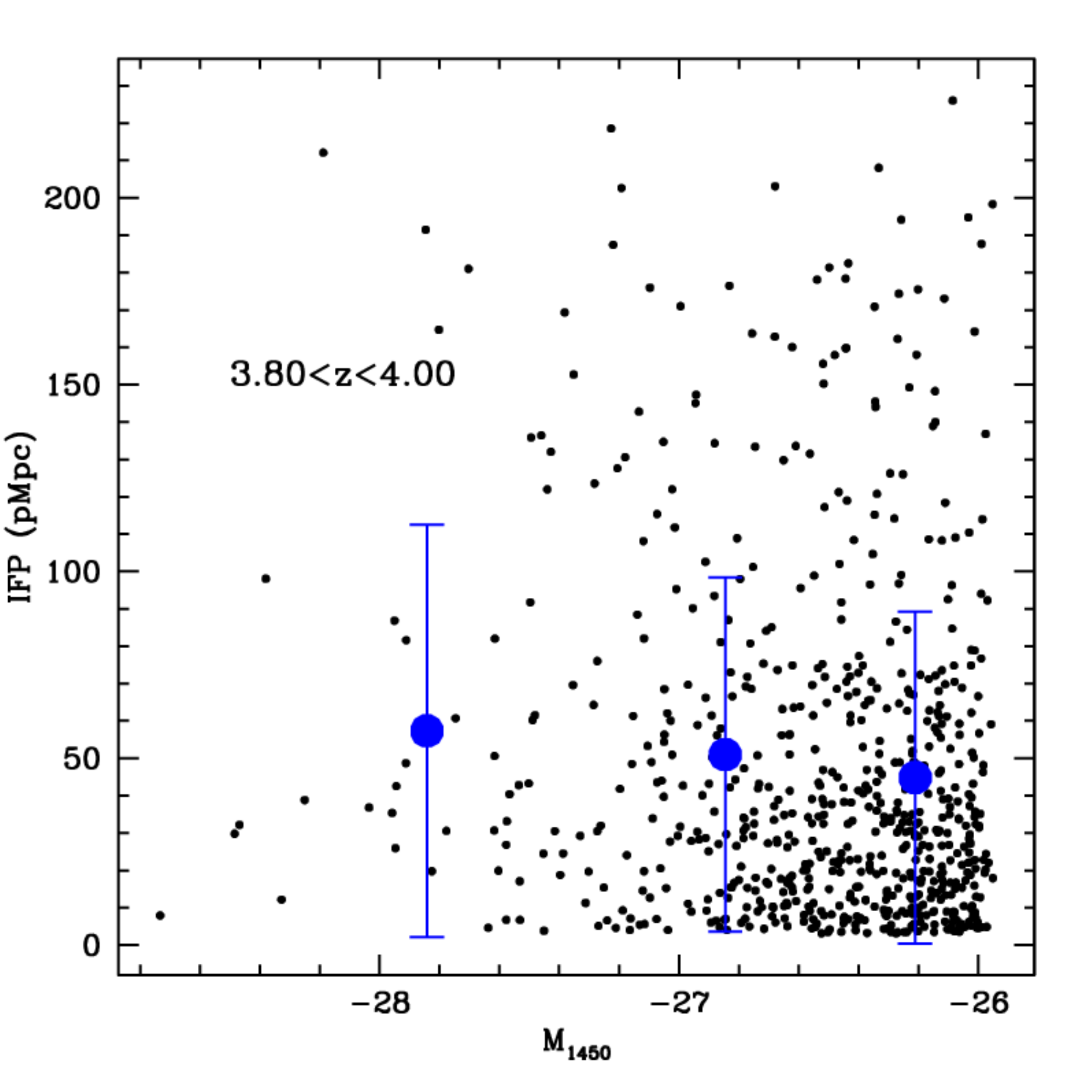}
\caption{
The small black dots represent the
free path vs absolute magnitude $M_{1450}$ for individual SDSS QSOs
at $3.8 < z < 4.0$ and $I\le 20.0$. The big blue points show the average value
of the free paths in bin of $\Delta M_{1450}=1.0$.
The error bars of the big blue points show the 84th and 16th percentiles of the individual free path distributions for each bin of absolute magnitude.
}
\label{MFPl}
\end{figure}

Fig. \ref{MFPl} shows the dependence of the IFP from the absolute
magnitude $M_{1450}$ of SDSS QSOs at $3.8 < z < 4.0$ and $I\le 20.0$. The
small black dots represent the measurements for individual QSOs, while
the big blue circles indicate the mean values of the free paths and absolute
magnitudes when dividing the sample in bins of $\sim$ 1.0 in $M_{1450}$. A
trend of larger IFP for brighter sources is possibly suggested by this plot,
with IFP $\sim 60$ pMpc at $M_{1450}\sim -28$ and IFP $\sim 45$ pMpc at
$M_{1450}\sim -26$. This trend is probably due to sources with low IFP
which are more frequent at fainter magnitudes (in the lower right
corner of Fig. \ref{MFPl}), rather than the lack of faint QSOs with long free paths.

\begin{figure}
\centering
\includegraphics[width=\columnwidth,angle=0]{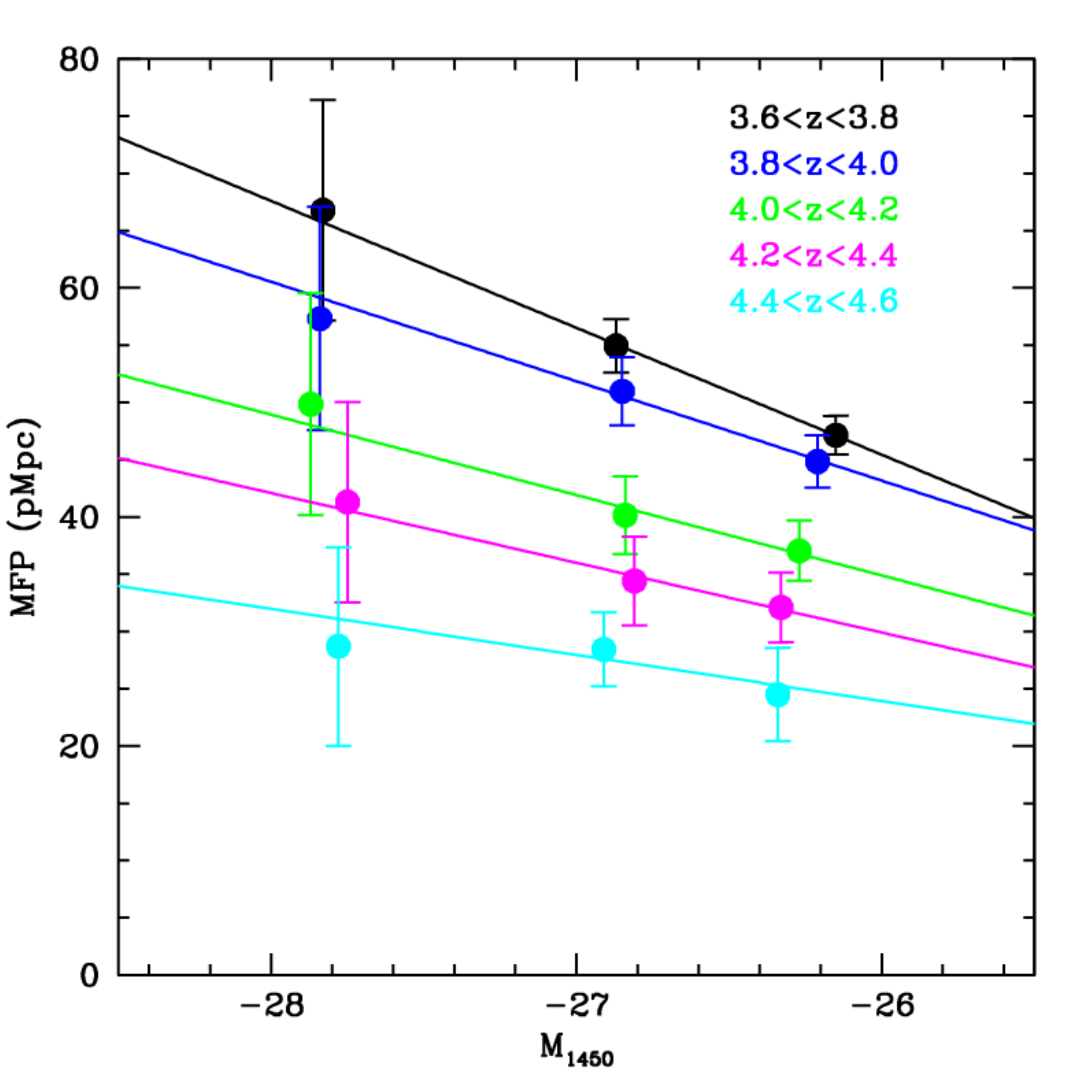}
\caption{
The average values of the free paths for all SDSS QSOs
at $3.6\le z\le 4.6$ and $I\le 20.0$, divided in redshift and absolute
magnitude $M_{1450}$ bins. The vertical bars are the
standard error of the mean in each bin of absolute magnitude.
The lines show the best linear fit to the observed mean
points for each redshift bin.
}
\label{MFPmag}
\end{figure}

A similar trend has been found also at higher redshifts, at least up
to $z=4.6$, the maximum redshift for our sample. Fig. \ref{MFPmag}
shows the mean values of the free paths for all the SDSS QSOs with $3.6\le z\le
4.6$ and $I\le 20.0$, divided in redshift and absolute magnitude
$M_{1450}$ bins. The continuous lines are the best fit with a linear
relation to the mean values. The trend with luminosity becomes milder
going at higher redshifts, and at $z\sim 4.5$ is almost
negligible.

\begin{figure}
\centering
\includegraphics[width=\columnwidth,angle=0]{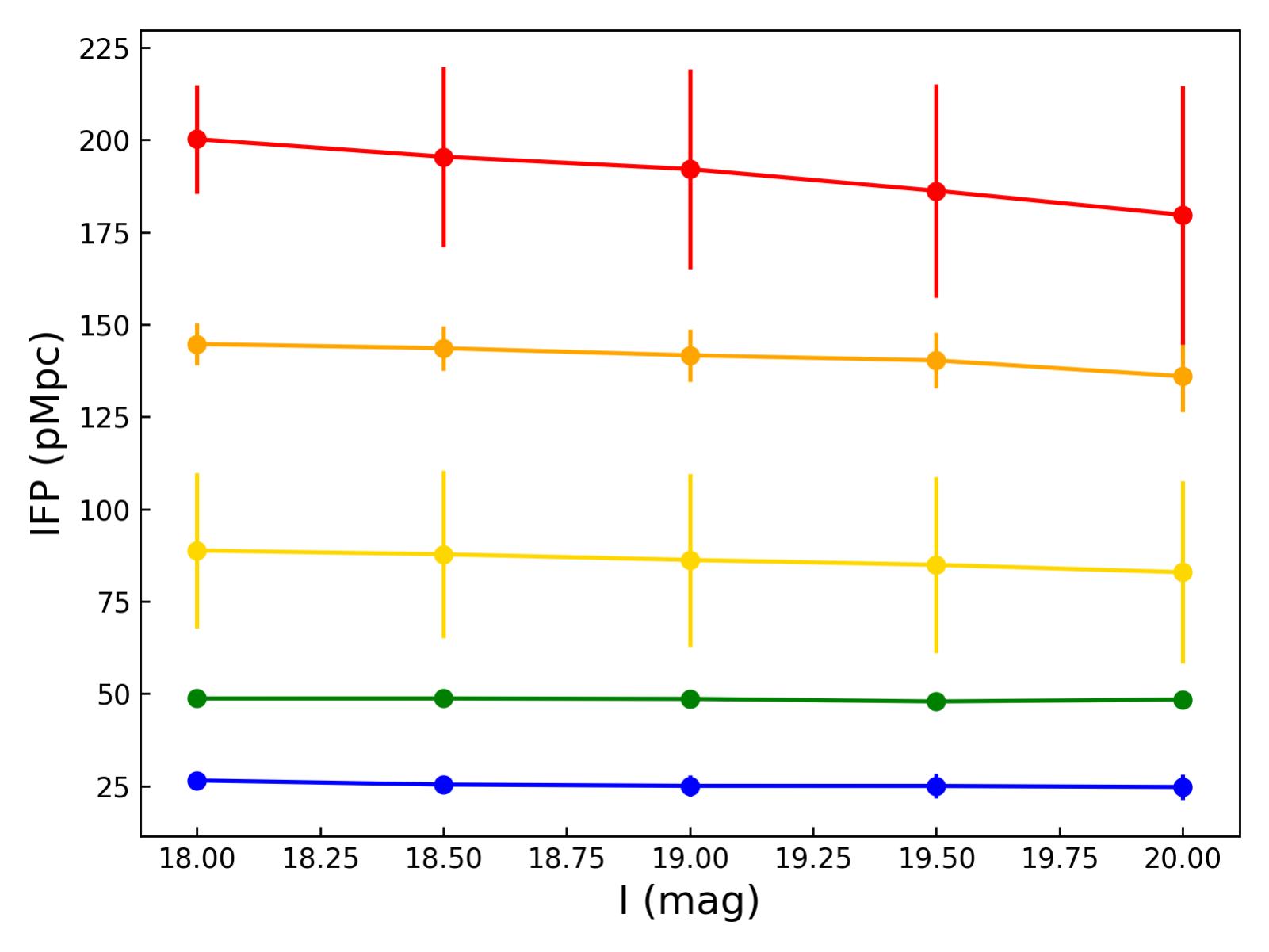}
\caption{
The free path of simulated QSOs at different I-band magnitudes. Circles show the mean value of 100 simulations, for each QSO and magnitude bin, while the error bars
indicate the 16th and 84th percentiles of the resulting IFP distribution. No strong trend of the IFP with the observed I-band magnitude has been detected.
}
\label{simmfp}
\end{figure}

While a redshift evolution of the mean free path is expected both by
theoretical and observational arguments, as discussed above,
however, an evolution in magnitude of the MFP is not so obvious.
In principle, we should expect that the physical properties of the IGM
(e.g., the MFP) will be independent from the properties of the tools
adopted to study it, i.e., the high-z QSOs analyzed in this paper. We
have carried out a simple simulation in order to check whether the
dependence of the IFP from the QSO luminosity can be due instead to an
instrumental effect, e.g., to the decreasing signal-to-noise ratio of the spectra.
We started from the observed spectra of five SDSS QSOs at $I\le 17.5$, i.e., with
high S/N ratio. We select them in order to span as much as possible the observed range of IFPs, from 20 to 200 pMpc.
These spectra have relatively high S/N and we
can measure their IFP with very high accuracy.
We used these spectra as input to our simulations:
for each simulated I-band magnitude in the range $I=18-20$,
we rescale the input spectra by the factor $10^{0.4(I_{obs}-I_{sim})}$, where $I_{obs}$ and $I_{sim}$ are the observed and simulated I-band magnitudes, respectively.
Then we add random
noise considering the original (i.e., without any rescaling) r.m.s. of the observed spectra released
by SDSS and derive the IFP of the simulated QSOs with the same tools
used for our measurements, as described above. We repeat this operation for 100 times for each simulated magnitude.
In Fig. \ref{simmfp} we plot the mean values of the output IFP for
each of the five QSOs and for each simulated magnitude
in the range $I_{sim}=18-20$, corresponding to the magnitude range of our sample. The error bars indicate the 16th and 84th percentiles of the IFP distribution.
We find no dependence of the
resulting IFP on the luminosity of the objects. This reassures us
that the observed trend with luminosity shown in Fig. \ref{MFPl} is not
spurious and can not come from the low S/N of the data at $M_{1450}\sim -26$.
Only at very large IFP ($\sim 200$ pMpc) and at $I\sim 20$ there is a slightly underestimate of the free path, of the order of $\sim 0.1$ dex. This effect however is too small to explain the strong trend found in Fig. \ref{MFPl}.
Moreover, it is not expected that at faint observed magnitudes
the SDSS algorithms are selecting preferentially sources with lower
free paths w.r.t. bright QSOs.
It is not so obvious to find a plausible explanation for the observed trend with luminosity.

\begin{figure}
\centering
\includegraphics[width=\columnwidth,angle=0]{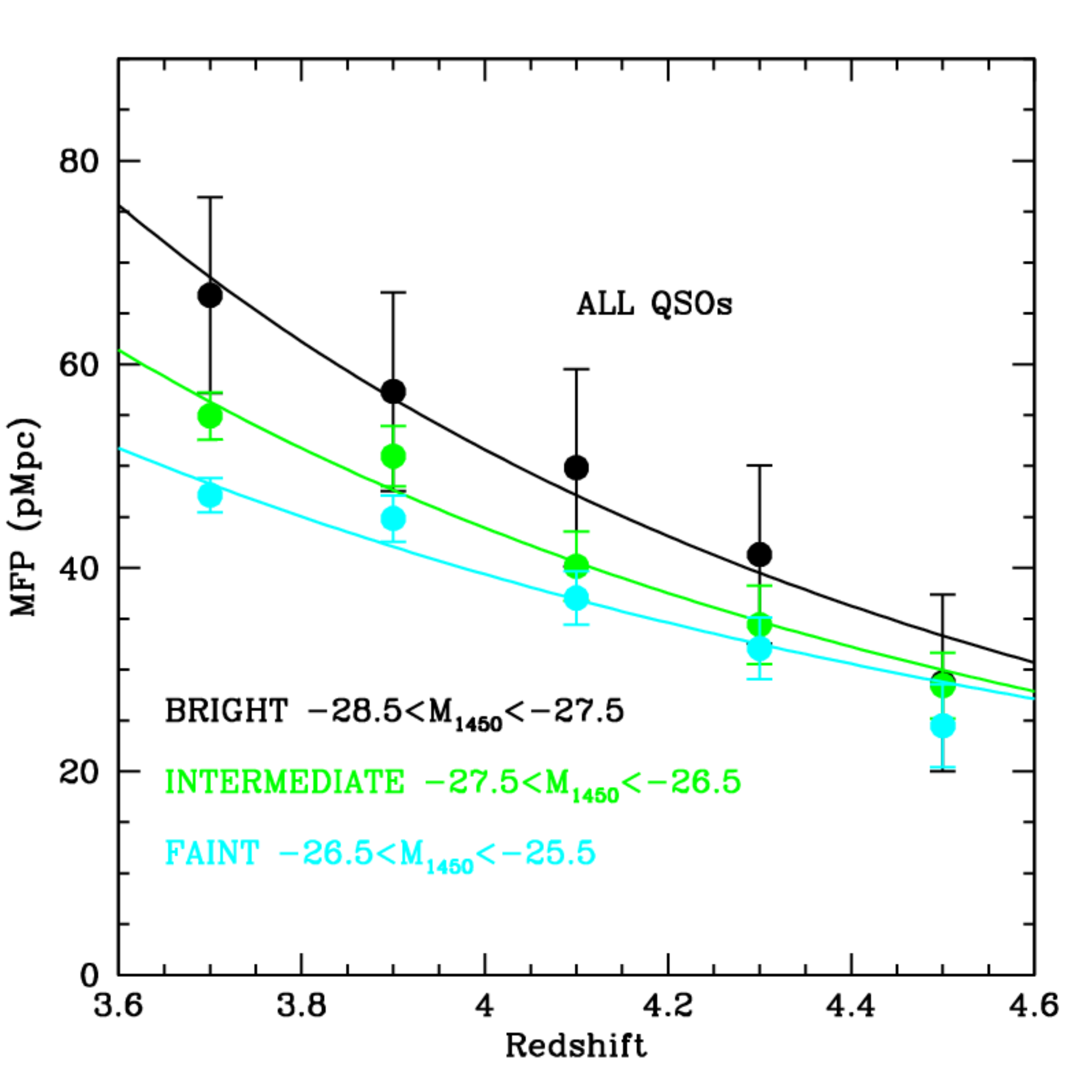}
\caption{
The mean values of the IFP for SDSS QSOs
at $3.6\le z\le 4.6$ and $I\le 20.0$ ("option 1"), divided in redshift for three bins of
absolute magnitude $M_{1450}$.
The vertical bars are the
standard error of the mean in each bin of absolute magnitude.
The continuous lines show the best power-law fit to the observed mean
points for each $M_{1450}$ bin.
}
\label{MFPz}
\end{figure}

Fig. \ref{MFPz} summarizes the mean values of the IFP for three
different bins of absolute magnitudes of SDSS QSOs from $z=3.6$ to
$z=4.6$ and magnitude $I\le 20.0$. The mean values of IFP have been
fitted with a power-law relation MFP$(z)$=MFP$(z=4)\times [(1+z)/5]^\eta$ and
the continuous lines show the best fit at different luminosities.
The redshift evolution $\eta$ we are finding is milder than the one by \citet{worseck14}. If we exclude from the fit the redshift interval between $z=3.6$ and 3.8 (we will see in the next sections that this bin could be prone to a systematic effect), the resulting $\eta$ is consistent with
the one by \citet{worseck14}, as shown in the bottom part of Table \ref{tabmfzall}.
It should be noted also that the redshift range studied by \citet{worseck14} is $2.3<z<5.5$, while our redshift interval is smaller ($3.6<z<4.6$) and thus could be more prone to uncertainties in the derivation of $\eta$.

\begin{table}
\caption{Redshift evolution of the MFP for all QSOs at $3.6 \leq z \leq 4.6$ and $I\le 20.0$ ("option 1"). The first two columns are the best fit slope and normalization at $z=4$, respectively. The last column indicates the absolute magnitude bins.
For comparison, the best fit values by \citet{worseck14} are MFP$(z=4)=37\pm2$ pMpc and $\eta=-5.4\pm0.4$.
In the bottom part,
we have excluded the $3.6<z<3.8$ redshift bin from the fit of the
MFP$(z)$ relation, since it could be affected by a possible underestimate
of the free path due to the limited wavelength coverage of the SDSS spectra in the bluer
wavelengths.
}
\label{tabmfzall}
\centering
\begin{tabular}{c c c}
\hline
\hline
$\eta$ & MFP$(z=4)$ & Magnitude bin $M_{1450}$ \\
 & pMpc & \\
\hline
\multicolumn{3}{c}{All redshift bins} \\
-4.6 $\pm$ 0.6 & 51.6 $\pm$ 1.6 & $-28.5\le M_{1450}\le -27.5$ \\
-4.0 $\pm$ 0.5 & 43.9 $\pm$ 1.1 & $-27.5\le M_{1450}\le -26.5$ \\
-3.3 $\pm$ 0.6 & 39.4 $\pm$ 1.2 & $-26.5\le M_{1450}\le -25.5$ \\
\hline
\multicolumn{3}{c}{Excluding 3.6 - 3.8 redshift bin} \\
-5.3 $\pm$ 1.0 & 53.2 $\pm$ 2.4 & $-28.5\le M_{1450}\le -27.5$ \\
-5.1 $\pm$ 0.3 & 45.7 $\pm$ 0.5 & $-27.5\le M_{1450}\le -26.5$ \\
-4.7 $\pm$ 0.4 & 40.8 $\pm$ 0.6 & $-26.5\le M_{1450}\le -25.5$ \\
\hline
\hline
\end{tabular}
\\
\end{table}

Interestingly, from Table \ref{tabmfzall} it is shown that, for our
fainter absolute magnitude bin $-26.5\le M_{1450}\le -25.5$, the best
fit to our mean values has a normalization MFP$(z=4)=39.4$ pMpc, which
is quite close to the best fit value of 37 pMpc provided by \citet{worseck14}.
This is not surprising, given the fact that their results are expected
to be dominated by faint QSOs with $M_{1450}\sim -26$, which is close to the
completeness limit of SDSS survey at $z\sim 4$.
The MFPs measured both by \citet{prochaska09} and by \citet{worseck14}
are based indeed on a stack of 150 QSOs for each redshift bin
adopted, with a bootstrap on the whole SDSS sample. Since these 150
QSOs are chosen randomly from the whole SDSS QSO survey in the adopted
redshift bin, the sample is possibly dominated by fainter QSOs, and
thus with lower IFPs, if the observed IFP-Luminosity relation turns out to be correct. In order to deepen the observed correlation between the IFP and the luminosity, we have cut our sample to IFP > 30 pMpc, finding that the luminosity dependency of IFP disappears. This reassures us that this trend is due to QSOs with low values of IFP, mainly at low luminosities, which could be possibly affected by
associated absorbers. However, we can not exclude that this effect is due to some biases caused by color selection effects which depend on the
observed magnitudes. A further investigation is needed in the future to completely explain the observed correlation.

In order to be sure that the examined QSO sample did not present too
much different features from that of \citet{prochaska09} or \citet{worseck14}, the stacking technique
was also implemented to infer the evolution of the mean free path in
redshift. The mean free path was computed from the rest-frame stacked
spectra in different redshift bins of $\Delta z=0.2$ from $z=3.6$ to
$z=4.6$, modeling the observed flux in the region blue-ward of the
Lyman limit as $f=f_{912}$ $\mathrm{exp}(-\tau_\mathrm{eff,LL})$, taking into account the spectral slope $\alpha_{\nu}$
and the softening of the QSO spectrum at $\lambda_{rest}\lesssim 1000$ by \citet{cristiani16}.
In particular, the stacked
spectra were averaged without weighting, in order not to introduce biases caused
by sources without strong LLS absorption, which have higher S/N ratio \citep{prochaska09}. The MFP
from the stack spectra were obtained by calculating the $\lambda_\mathrm{LLS}$
at which $\tau_\mathrm{eff,LL}=1$, as described in section \ref{subsec:mfp_estimate}, and converting this wavelength into a
redshift $z_\mathrm{LLS}$ by assuming a rest-frame $\lambda_{rest}=912$ {\AA}.

As discussed in Section \ref{subsec:mfp_estimate} (and in the forward in Section \ref{subsub:limited_cov}), the
MFP estimate for QSOs at $3.6<z<3.8$ could be biased low due to the
limited coverage of the observed SDSS spectra ($\lambda_{obs}\ge 3560$
{\AA}). We have thus excluded the redshift bin $3.6<z<3.8$ from the
fit in Fig. \ref{MFPmag} and obtained steeper evolution of the MFP with
redshift (Table \ref{tabmfzall}, bottom part). As shown by our results,
the redshift evolution is still milder than the one by \citet{worseck14}.

\begin{figure}
\centering
\includegraphics[width=\columnwidth,angle=0]{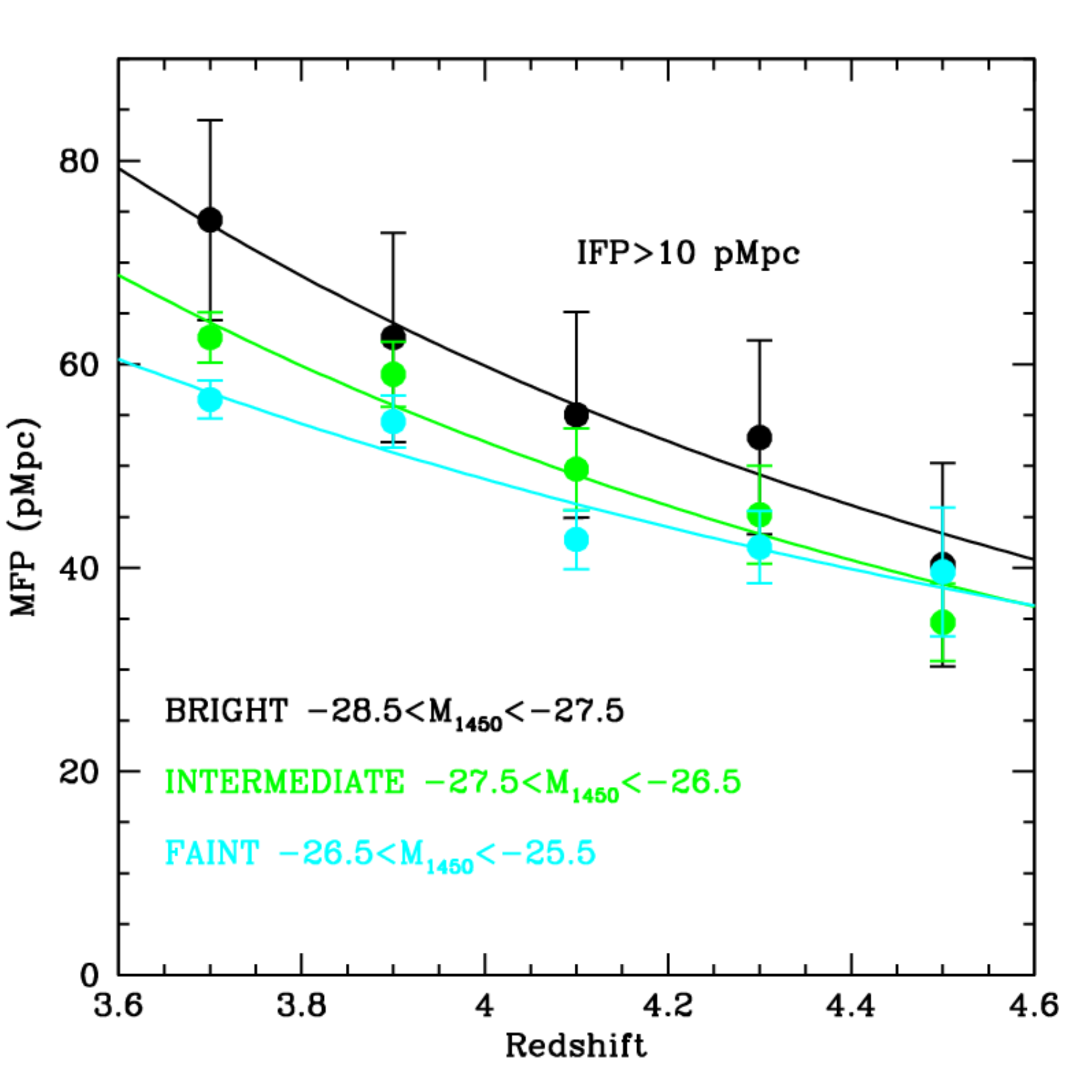}
\caption{
The mean values of the IFP for SDSS QSOs
at $3.6\le z\le 4.6$, $I\le 20.0$, and IFP $> 10$ pMpc ("option 2"),
divided in redshift for three bins of
absolute magnitude $M_{1450}$.
The vertical bars are the
standard error of the mean in each bin of absolute magnitude.
The continuous lines show the best power-law fit to the observed mean
points for each $M_{1450}$ bin.
}
\label{MFPmagnoaa}
\end{figure}

If we take out from our sample all the QSOs with IFP $\le 10$ pMpc, assuming that they are affected by
associated absorptions ("option 2"), the trends highlighted above are further migrating towards a
milder evolution of MFP with redshift. Fig. \ref{MFPmagnoaa} shows the
updated mean values of IFP, divided in different bins of absolute magnitude,
versus redshift.
It is immediately evident that the mean values of
the IFP are significantly higher than the \citet{worseck14} best fit
(MFP $\sim 50$ pMpc at $z=4$ vs MFP $=37\pm 2$ pMpc by \cite{worseck14})
and the redshift evolution is much milder. Table \ref{tabmfznoaa}
summarizes the best-fit values of our mean free path with a power law
relation. At all luminosities the slope is $\eta\sim -2.6\div -4.3$, which
is significantly milder than the \citet{worseck14} one ($\eta=-5.4\pm 0.4$).

\begin{table}
\caption{Redshift evolution of the mean free path for QSOs at $3.6 \leq z \leq 4.6$
with IFP $>$ 10 pMpc ("option 2"). We have excluded QSOs with IFP $\le 10$ pMpc (possibly affected by associated
absorbers) from the fit of the MFP$(z)$ relation.
In the bottom part,
we have excluded the $3.6<z<3.8$ redshift bin from the fit of the
MFP$(z)$ relation, since it could be affected by a possible underestimate
of the IFP due to the limited wavelength coverage of the SDSS spectra in the bluer wavelengths.}
\label{tabmfznoaa}
\centering
\begin{tabular}{c c c}
\hline
\hline
$\eta$ & MFP$(z=4)$ & Magnitude bin $M_{1450}$ \\
 & pMpc & \\
\hline
\multicolumn{3}{c}{All redshift bins} \\
-3.4 $\pm$ 0.4 & 59.8 $\pm$ 1.4 & $-28.5\le M_{1450}\le -27.5$ \\
-3.3 $\pm$ 0.5 & 52.4 $\pm$ 1.4 & $-27.5\le M_{1450}\le -26.5$ \\
-2.6 $\pm$ 0.6 & 48.7 $\pm$ 1.4 & $-26.5\le M_{1450}\le -25.5$ \\
\hline
\multicolumn{3}{c}{Excluding 3.6 - 3.8 redshift bin} \\
-3.2 $\pm$ 0.8 & 59.4 $\pm$ 2.5 & $-28.5\le M_{1450}\le -27.5$ \\
-4.3 $\pm$ 0.5 & 54.4 $\pm$ 1.2 & $-27.5\le M_{1450}\le -26.5$ \\
-3.3 $\pm$ 1.0 & 49.7 $\pm$ 1.8 & $-26.5\le M_{1450}\le -25.5$ \\
\hline
\hline
\end{tabular}
\\
\end{table}

We have also checked the effect of eliminating BAL QSOs from our sample. First of all, it is worth noting that the analysis carried out by \citet{worseck14} could include sources with possible associated absorbers, but excludes all the BAL QSOs. We have found
that both the normalization and the slope of the MFP$(z)$ relation are
not affected by the presence of BAL QSOs, and the best fit values of
the sample without BAL QSOs (and without AAs) are very
similar to the ones in Table \ref{tabmfznoaa}. This effect is not
surprising, since both the LyC escape fraction and the individual free path
distributions do not depend on the BAL QSO fraction. This indicates
implicitly that there is no apparent relation from the BAL
phenomenon and the associated absorbers.

\begin{figure}
\centering
\includegraphics[width=\columnwidth,angle=0]{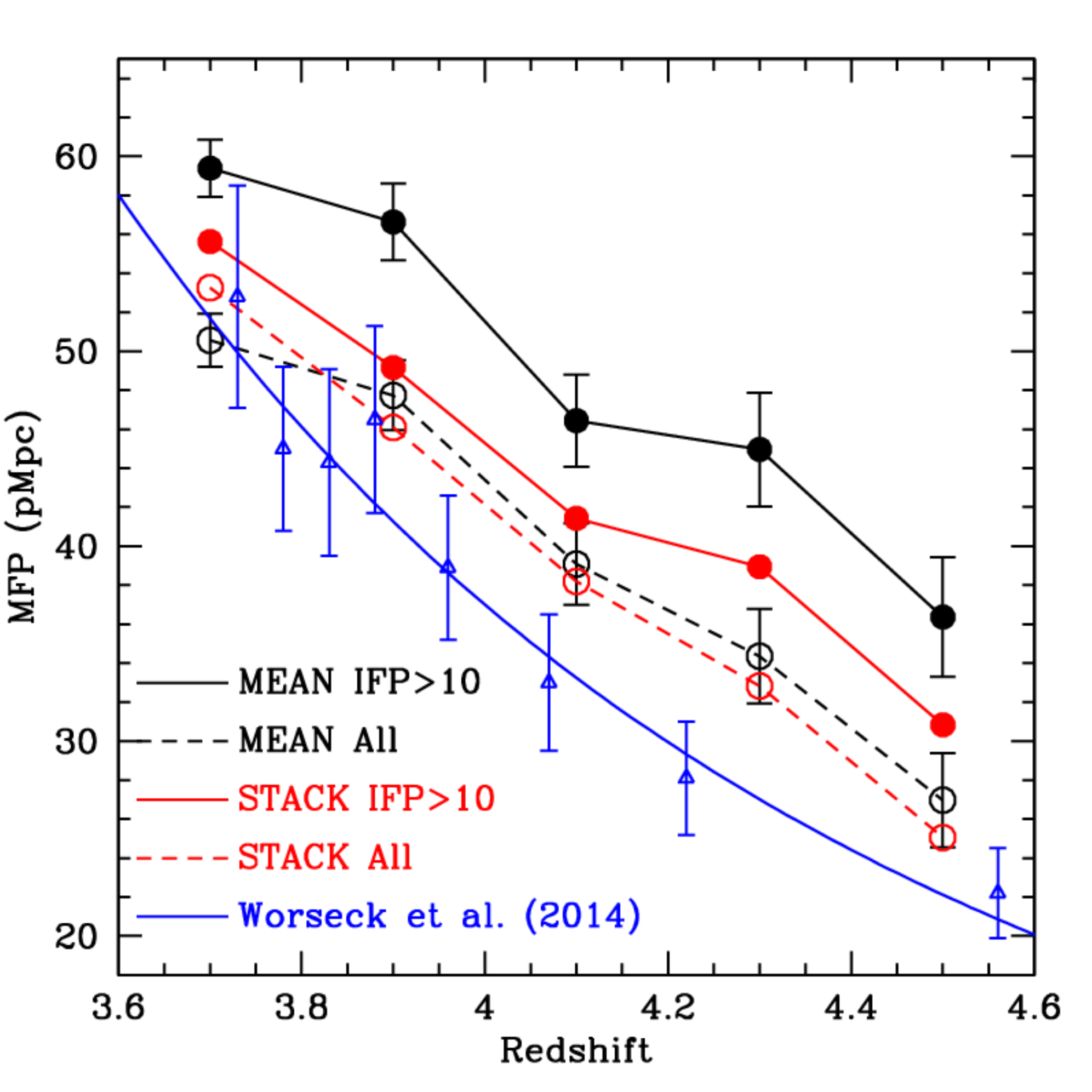}
\caption{
The redshift dependence of the MFP for SDSS QSOs
at $3.6\le z\le 4.6$ and $I\le 20.0$.
The filled black points (connected by a continuous black line)
summarize the mean value of the free paths excluding objects
with IFP $\le 10$ pMpc, which are possibly affected by associated absorbers ("option 2").
The open black circles (connected by a dashed black line) show the MFP
for all the QSOs of our sample ("option 1"). The filled red points (connected by a
continuous red line) instead are derived from the stack of QSOs with
IFP $> 10$ pMpc, and the open red circles (connected by a dashed red line)
show the MFP obtained by the stack of all the QSOs.
The dotted blue curve is the evolution of the MFP with redshift found by
Worseck et al. (2014), while the blue open triangles indicate their mean values in different redshift bins.
}
\label{summaryMFP}
\end{figure}

Fig. \ref{summaryMFP} summarizes the main results of this section. If
we consider all the QSOs at $3.6 \leq z \leq 4.6$ and $I\le 20.0$ ("option 1"), the mean
value of the PDF(IFP) and the stack give similar results (as we discuss in Appendix B), while if we
exclude QSOs with IFP $\le 10$ pMpc, assuming they are affected by associated absorbers ("option 2"),
we obtain a huge difference between mean and stack and a milder
redshift evolution w.r.t. \citet{worseck14}. Table \ref{tabsummaryMFP}
summarizes the MFP obtained through stack and mean values of the
PDF(IFP), compared with the results of \citet{worseck14}.

\begin{table*}
\caption{Summary of the MFP vs redshift. In the first column are the redshift bins. MFP$_\mathrm{W14}$ indicates the MFP from the best-fit formula by \citet{worseck14}. MFP$\mathrm{^{all}_{stack}}$ and MFP$\mathrm{^{all}_{mean}}$ are the values obtained by the stack technique and the mean of the distribution functions, respectively, for the whole sample ("option 1"). MFP$\mathrm{^{noAAs}_{stack}}$ and MFP$\mathrm{^{noAAs}_{mean}}$ are the same quantities excluding QSOs with IFP $\le$ 10 pMpc ("option 2"). The last two columns are the ratio between the MFP values obtained by the distribution functions for "option 1" and "option 2", respectively, and MFP$_\mathrm{W14}$.}
\label{tabsummaryMFP}
\centering
\begin{tabular}{l c c c c c | c c}
\hline
\hline
Redshift bin & MFP$_\mathrm{W14}$ & MFP$\mathrm{_{stack}^{all}}$ & MFP$\mathrm{_{mean}^{all}}$ &
MFP$\mathrm{_{stack}^{noAAs}}$ & MFP$\mathrm{_{mean}^{noAAs}}$ &
$\frac{\mathrm{MFP}_\mathrm{op1}}{\mathrm{MFP}_\mathrm{W14}}$ &
$\frac{\mathrm{MFP}_\mathrm{op2}}{\mathrm{MFP}_\mathrm{W14}}$ \\
 & pMpc & pMpc & pMpc & pMpc & pMpc & &\\
\hline
3.60-3.80 & 51.68 & 53.26 & 50.56 & 55.63 & 59.39 & 0.978 & 1.149 \\
3.80-4.00 & 41.27 & 46.08 & 47.74 & 49.17 & 56.64 & 1.157 & 1.373 \\
4.00-4.20 & 33.25 & 38.20 & 39.09 & 41.43 & 46.44 & 1.176 & 1.397 \\
4.20-4.40 & 27.01 & 32.83 & 34.36 & 38.95 & 44.96 & 1.272 & 1.665 \\
4.40-4.60 & 22.11 & 25.06 & 26.97 & 30.83 & 36.37 & 1.220 & 1.645 \\
\hline
\hline
\end{tabular}
\\
\end{table*}

The differences between our best fit and the
\citet{worseck14} results could be due to two main reasons: 1-we
exclude from our sample the QSOs which are suspected to be affected by
associated absorbers, with IFP $\le 10$ pMpc ("option 2"); 2-we have used the mean
values of the PDF(IFP), instead of the stack as done by
\citet{prochaska09} and \citet{worseck14}. As discussed in the
Appendix \ref{appendix:B}, an underestimate of the MFP from stacking compared to the
mean value of the PDF is expected, in particular for skewed
distributions as we are finding at $z\gtrsim 4$.
Following these arguments, the probability distribution function of individual free path of UV ionizing photons can give more information on the IGM than the stacking technique. For this reason, in the following we will use the PDF(IFP) in order to compute the contribution of QSOs/AGNs to the HI ionizing background.

From Fig. \ref{MFPmagnoaa} it is interesting also to note that the
difference between the bright and faint QSO samples is reducing at
$z>4.2$ and it seems almost negligible at $z\sim 4.5$.
Moreover, the mean value of the IFP distributions is
progressively diverging from the \citet{worseck14} estimates from $z=3.6$ to $z=4.6$ for "option 2".
This could have
deep implications for the MFP estimate at the epoch of reionization. For
example, from \citet{worseck14} relation a mean free path of $\sim 6$
pMpc is expected at $z=6$, while a MFP three times larger (MFP $\sim
18$ pMpc) is awaited extrapolating our relation in Table
\ref{tabsummaryMFP} and Fig.\ref{summaryMFP}. If the outlined trend
shown in Fig.\ref{MFPmagnoaa} for "option 2" is kept also at higher redshifts, this
could have important implications for the computation of the
photo-ionization rate at the EoR, as we will discuss in the next
sub-section.

\subsection{The corrected ionizing background at z=4 and implications on Reionization}

The contribution of bright QSOs, faint AGNs and/or star-forming galaxies to the HI ionizing
background can be quantified by computing the photo-ionization rate of
neutral hydrogen atoms, $\Gamma_\mathrm{HI}$. This parameter primarily
depends on the overall emissivity of the sources and on their
mean free path, i.e., $\Gamma_\mathrm{HI}\propto \mathrm{MFP}\cdot f_{esc}\cdot \epsilon_{912}$,
where $f_{esc}$ is the escape fraction of Lyman continuum
photons and $\epsilon_{912}$ is the overall intrinsic contribution of the ionizing sources
emitted at 912 {\AA} rest-frame. The latter depends on the luminosity
function of high-z AGNs or SFGs and on the spectral slope of the sources close to
912 {\AA} rest-frame.
Since it is not the aim of this paper to revise the estimate
of the emissivity of the different populations of galaxies and AGNs, we will only compute here the
scaling factor due to our revised estimates of the MFP and $f_{esc}$ at $z\sim 4$,
$\Gamma_\mathrm{HI}^\mathrm{new}/\Gamma_\mathrm{HI}^\mathrm{old}$.

If the main sources of HI ionizing photons are star-forming galaxies at high redshift, the scaling factor for $\Gamma_\mathrm{HI}$ is simply the ratio MFP$_\mathrm{mean}$/MFP$_\mathrm{W14}$, from Table \ref{tabratio}. It varies from 0.98 to 1.27, if all the QSOs are used for an estimate of the mean free path ("option 1"). The correction is significantly higher if the mean free path is derived only by QSOs with IFP $>$ 10 pMpc ("option 2"). In the latter case, the ratio
MFP$_\mathrm{mean}/$MFP$_\mathrm{W14}$ increases from $\sim$ 1.1 at $z=3.7$ to 1.6 at $z=4.5$.

In the opposite case that QSOs and AGNs are the main producers of the ionizing background at $z\gtrsim 4$,
the ratio $\Gamma_\mathrm{HI}^\mathrm{new}/\Gamma_\mathrm{HI}^\mathrm{old}$ does not
simply scale as MFP$_\mathrm{mean}$/MFP$_\mathrm{W14}$.
We should take into account two additional correction factors, i.e., 1-the possible correlation between individual free path and escape fraction (see Fig.\ref{fescMFP}) and 2-the fraction of QSOs with IFP greater than 10 pMpc for "option 2".
The first factor ($CORR1$ in Table \ref{tabratio}) has been computed as $<\mathrm{IFP}*f_{esc}>/(<\mathrm{IFP}>*<f_{esc}>)$ and is always greater than 1, given the observed correlation between these two parameters (see Fig.\ref{fescMFP}). The fraction $CORR2$ of QSOs with IFP $> 10$ pMpc, instead, is of the order of 0.7-0.8 for "option 2", while $CORR2$ is always 1.0 for "option 1", by definition.

If we consider all the QSOs ("option 1"), then the new estimate of MFP$_\mathrm{mean}$ is not significantly larger than previous estimates, but this is balanced by a strong correction for the correlation between IFP and escape fraction ($CORR1\sim 1.2-1.3$).
In this case the global correction is of the order of $\sim 1.2-1.7$ (upper part of Table \ref{tabratio}).

If we instead exclude from our calculations QSOs with IFP $\le 10$ pMpc ("option 2"), the higher value of MFP is compensated by the fact that only 70-80\% of the lines of sight are contributing to the photo-ionization rate. In this case (bottom part of Table \ref{tabratio}) the total correction to $\Gamma_\mathrm{HI}$ is 1.2 in the redshift interval $z=3.6-3.8$, rising to 1.4-1.7 at higher redshift.
In addition, this ratio increases from $z=3.7$ to $z=4.5$.
Interestingly, the global correction to the ionizing background $\Gamma_\mathrm{HI}^\mathrm{new}/\Gamma_\mathrm{HI}^\mathrm{old}$ does not depend whether QSOs with IFP $\le 10$ pMpc are included ("option 1") or excluded ("option 2") in our calculations, but it is practically consistent between the two cases.

\begin{table}
\caption{Total corrections to the HI photo-ionizing background. The redshift bins are indicated in the first column.
MFP$_\mathrm{mean}$/MFP$_\mathrm{W14}$ are the ratio between our value
MFP$_\mathrm{mean}$ and MFP$_\mathrm{W14}$, which are taken from Table \ref{tabsummaryMFP}.
$N_\mathrm{QSO}$ is the number of sources in each redshift bin
with IFP above a given threshold.
$CORR1$ is the correction term which takes into account the correlation between escape fraction and IFP, as observed in Fig.\ref{fescMFP}, and it is computed as $<\mathrm{IFP}*fesc>/(<\mathrm{IFP}>*<fesc>)$.
$CORR2$ is the fraction between the number of QSOs with IFP above a given threshold and the total number of QSOs in each redshift bin.
$\Gamma_\mathrm{HI}^\mathrm{new}/\Gamma_\mathrm{HI}^\mathrm{old}$ is the product of
MFP$_\mathrm{new}$/MFP$_\mathrm{W14}$ with $CORR1$ and $CORR2$.}
\label{tabratio}
\centering
\begin{tabular}{l c c c c c}
\hline
\hline
Redshift & $\frac{\mathrm{MFP}_\mathrm{mean}}{\mathrm{MFP}_\mathrm{W14}}$ & $N_\mathrm{QSO}$ & $CORR1$ & $CORR2$ & $\Gamma_\mathrm{HI}^\mathrm{new}/\Gamma_\mathrm{HI}^\mathrm{old}$ \\
bin & & & & & \\
\hline
\multicolumn{6}{c}{Option 1: all QSOs} \\
3.6-3.8 & 0.978 & 1099 & 1.283 & 1.000 & 1.255 \\
3.8-4.0 & 1.157 &  668 & 1.258 & 1.000 & 1.455 \\
4.0-4.2 & 1.176 &  382 & 1.314 & 1.000 & 1.545 \\
4.2-4.4 & 1.272 &  219 & 1.370 & 1.000 & 1.742 \\
4.4-4.6 & 1.220 &  140 & 1.402 & 1.000 & 1.710 \\
\hline
\multicolumn{6}{c}{Option 2: only QSOs with IFP $>10$ pMpc} \\
3.6-3.8 & 1.149 & 915 & 1.309 & 0.833 & 1.252 \\
3.8-4.0 & 1.373 & 550 & 1.283 & 0.823 & 1.450 \\
4.0-4.2 & 1.397 & 312 & 1.350 & 0.817 & 1.540 \\
4.2-4.4 & 1.665 & 159 & 1.428 & 0.726 & 1.726 \\
4.4-4.6 & 1.645 &  96 & 1.492 & 0.686 & 1.683 \\
\hline
\hline
\end{tabular}
\\
\end{table}

It is worth noting that the values of MFP$_\mathrm{mean}/$MFP$_\mathrm{W14}$ inferred in this work are only lower limits. In fact, the averaged values of the mean free path computed from the IFP distribution functions may be even greater if all the possible caveats in the estimate of this parameter are taken into account (i.e., observational limitations and selection effects of some QSOs, see Section \ref{sec:systematics}).

In summary, our results indicate a milder evolution of the MFP with redshift, giving a revised estimates of the photo-ionization rate by a factor of 1.4-1.7 upward at $z\sim 4.0-4.5$. If the trend obtained with these QSOs is extrapolated at $z\sim 6$, then the new estimates of the MFP, and consequently of the ionizing background, could be a factor of 2-3 times higher than the recent estimates, both for star-forming galaxies and QSOs/AGNs (e.g., \cite{finkelstein19,giallongo19}). In the future it will be important to check whether this trend is still present at $z>5$.


\section{Discussions}

\subsection{Possible systematic effects}\label{sec:systematics}

This work is based on a sample of $z\sim 4$ QSOs selected by the SDSS
thanks to their optical photometry \citep{paris18}. The specific
selection criteria adopted by SDSS may affect our results, especially
for what concerns the escape fraction and mean free path
estimates. We take into account here the effect of the limited
extension of SDSS spectra in the UV and the color selections of $z\sim
4$ QSOs. Both these issues could prevent a correct and robust
estimate of the total contribution of AGNs to the HI ionizing
background at $z\sim 4$. We do not attempt here to correct for these
two effects, but it is easy to conclude, from what we are going to
show in the following, that the net effect of these two issues leads to an underestimate of both the escape fraction and MFP. Dedicated observations and well-defined surveys are needed at the end to
properly correct for these problems.

\subsubsection{The limited coverage of SDSS spectra in the UV}\label{subsub:limited_cov}

The SDSS spectra released in the DR14 distribution are limited to the
wavelength range $3600\lesssim \lambda_{obs}\lesssim 10000$ {\AA}.
The cut at lower wavelengths in the UV can affect our estimate of
the free path of QSOs at $z\sim 4$, as shown in Fig. \ref{esMFP}. In this
case, this QSO (SDSS J113654.6+485322.3 at $z_\mathrm{spec}=3.61964$) has a long
path along the line of sight clear from damped Lyman-$\alpha$ or Lyman
limit systems. Its ionizing radiation is thus able to travel tens of
proper Mpc without being attenuated by a factor $1/e$ (defining the IFP)
and can reach the
observed lower limit of the SDSS spectrum at $\lambda_{obs}\sim 3550$
{\AA}. It is not known, with the present data, whether the ionizing
radiation of this QSO is able to reach bluer wavelengths, and for this
reason an individual free path of $145.94$ pMpc is assigned.
This is a
lower limit, simply imposed by an instrumental limitation; the real
value of the IFP for this object could be larger than the value assigned here.
In our sample, for 12.8\% of the QSOs the provided IFP is a lower limit due to this effect.
In the redshift bin $3.6<z<3.8$ this fraction is 15.4\% and it is progressively decreasing to
10\% in the highest redshift bins considered.
In the future it will be important to spectroscopically follow up these kind of objects to definitively measure their IFP and to ensure that the SDSS spectra are not affected by any systematics.

\begin{figure}
\centering
\includegraphics[width=\columnwidth,angle=0]{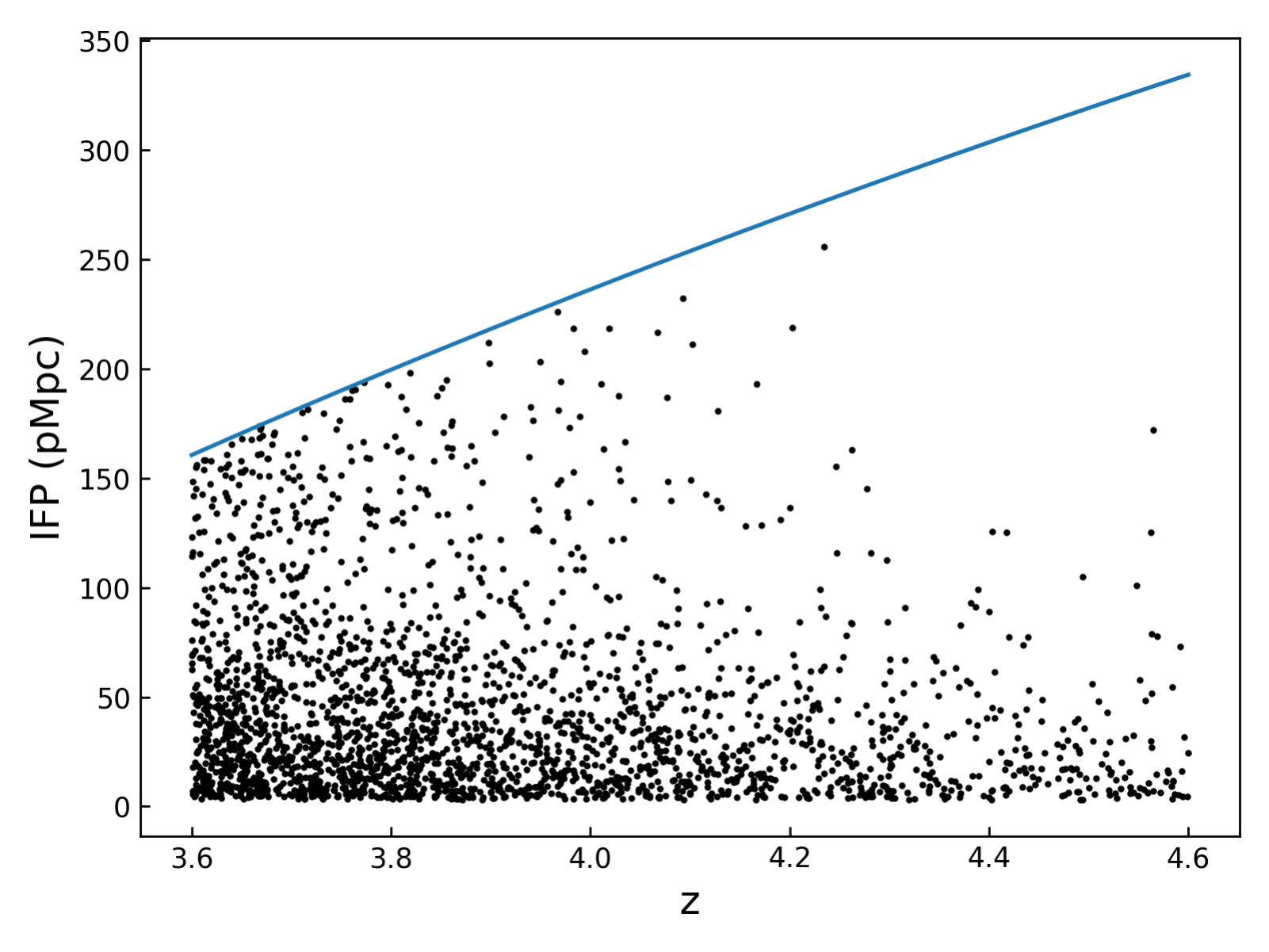}
\caption{
Individual free path (black points) vs spectroscopic redshift for SDSS QSOs
at $3.6\le z\le 4.6$. The blue line represents the maximum value of
IFP which can be obtained at each redshift due to the observational limits of the SDSS
spectra, i.e., $\lambda_{obs} \gtrsim 3550$ {\AA}.
}
\label{MFPredz}
\end{figure}

This issue causes the artificial lack of QSOs with long IFP at lower
redshifts, especially close to $z\sim 3.6$, as shown in
Fig. \ref{MFPredz}. Many sources with high free paths could present a
spectrum which extends far beyond the lower instrumental limit of the
observed wavelengths; this results in an altered estimate of the IFP
distribution. Since at $z\sim 4$ we are finding QSOs with IFP $\sim 200$
pMpc or larger, we do expect that these sources will be present also
at lower redshifts. Their absence however is due only to instrumental
limitation, as shown by Fig. \ref{MFPredz}. The blue
line in this plot represents the maximum values of the IFP that can be obtained
considering a minimum observable wavelength of about 3550 {\AA} at
different redshifts. These limits exactly follow the distributions of
the sources with the highest IFP, confirming that this trend is not a
physical effect, but an instrumental feature, as explained before. To
correct for this systematic effect the sources close to the
instrumental limit shown in Fig. \ref{MFPredz}, deep spectra with spectrographs
sensitive to UV bluer wavelengths are required, i.e., MODS1 at LBT or
FORS1 at VLT, as shown by \cite{grazian18}.

In conclusion, this particular issue can artificially reduce the average value and
the scatter of the free path obtained through the probability
distribution function of the QSO sample, especially at $z\le 3.9$. This effect is particularly noticeable in Fig. \ref{MFPz} and \ref{MFPmagnoaa}. For example, if we compute the MFP in the redshift interval $4.0\le z\le 4.1$, we obtain MFP $=40.73$ pMpc, while, if we exclude
sources with IFP $\ge 150$ pMpc (thus simulating the bias observed at $z\sim
3.6$), the average value is artificially lower, MFP $=34.61$
pMpc. Thus, the MFP adopted in the present paper at $z\le 3.9$ could
be higher, of the order of $\sim 18\%$.

This underestimate of the mean free path at $z\le 3.7$
can also artificially flatten the redshift evolution of this parameter shown in Fig.\ref{summaryMFP}. We have excluded the first redshift bin and compute the best fit for the sample of QSOs with IFP $> 10$ pMpc ("option 2"), and find $\eta\sim -4.3\div -3.2$, with slightly larger uncertainties ($\sigma_\eta\sim 0.8$). In the future,  deep spectra extending in the UV at $\lambda_{obs} \sim 3300$ {\AA} for these QSOs would provide a more robust estimate of the redshift evolution of the MFP.

\subsubsection{Color Selection Effects}

\begin{figure}
\centering
\includegraphics[width=\columnwidth,angle=0]{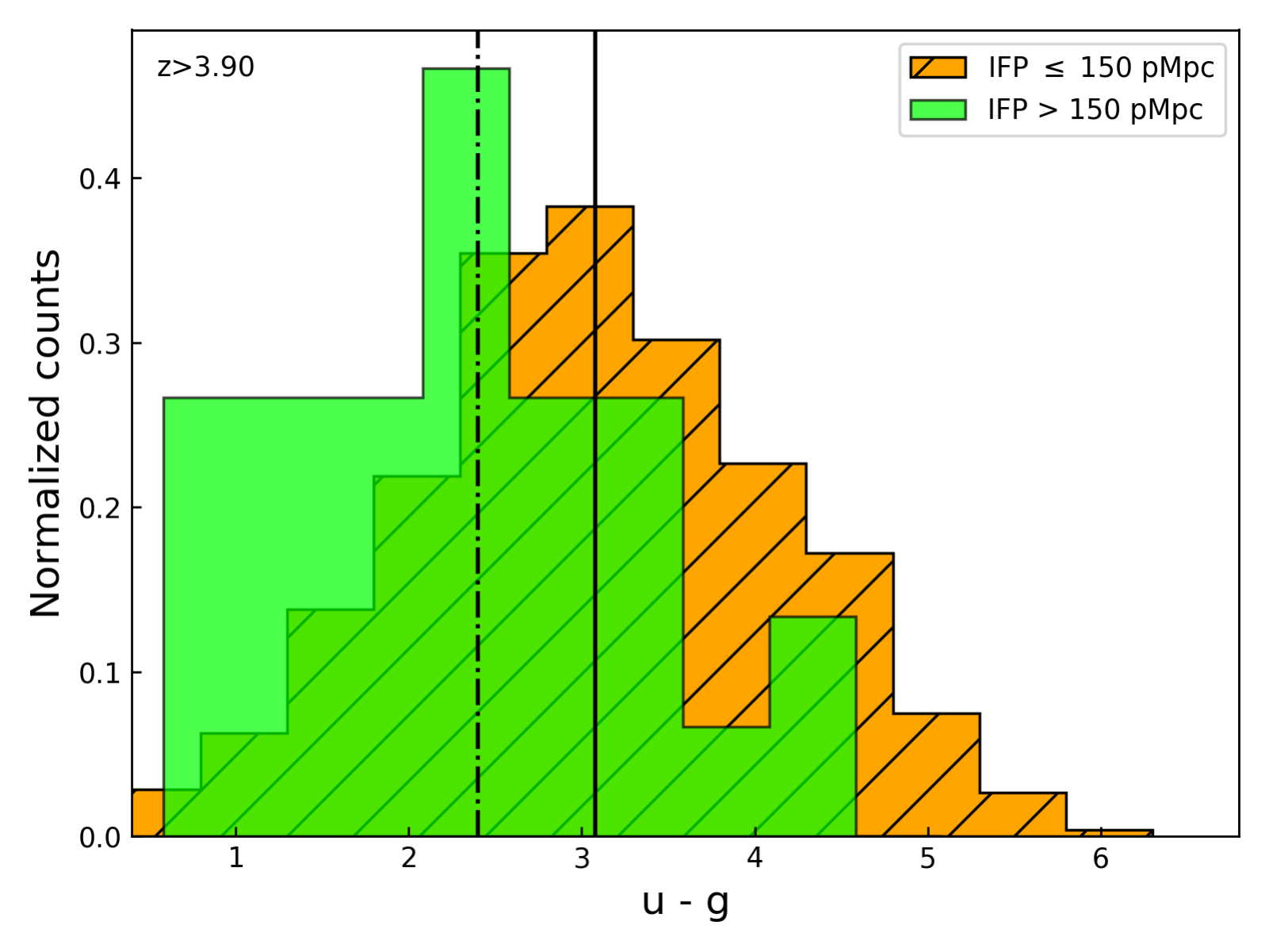}
\includegraphics[width=\columnwidth,angle=0]{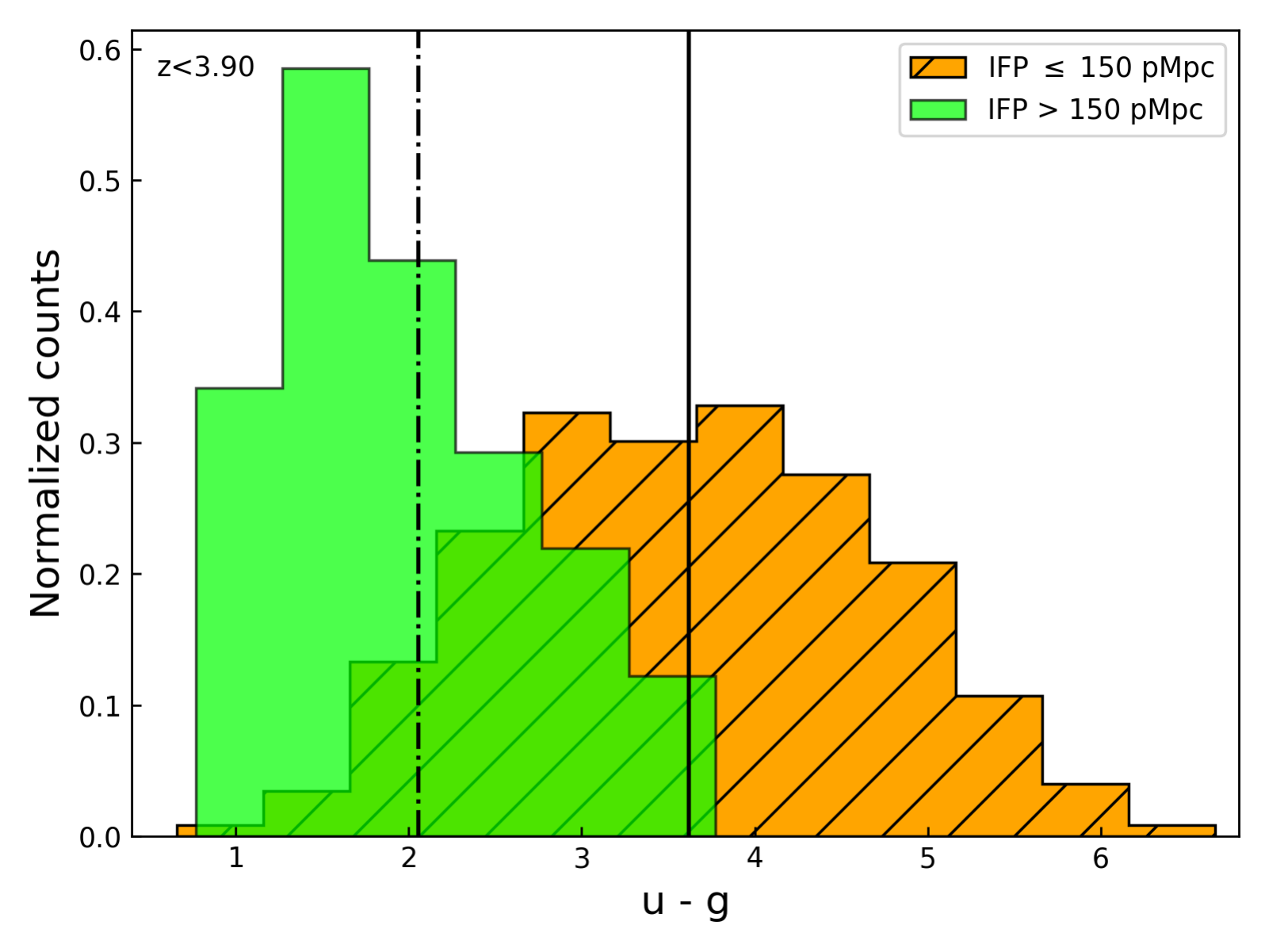}
\caption{Distribution functions of the $u-g$ color for QSOs with high (green histogram, IFP $\ge 150$ pMpc)
and low (orange hatched histogram, IFP $< 150$ pMpc) individual free paths. The solid and dot-dashed black lines refer to the average values
of the orange and green distributions, respectively. The top (bottom) panel shows the $u-g$ distributions for QSOs at higher (lower) redshift w.r.t. the median one at $z\sim3.90$.}
\label{ugMFP}
\end{figure}

The selection criteria adopted by the SDSS team to select the QSOs
analyzed in this paper are based on optical colors \citep{paris18}. In
particular, the selection of QSOs at $3.6\le z\le 4.2$ is mainly based
on the color criterion $u-g>1.5$ \citep{richards02}, which is especially useful because
it involves the Gunn filters $u$ and $g$, centered at the average
wavelengths 3551 and 4686 {\AA}, respectively. The $u-g$ color takes
advantage of the expected flux decrement of the source in the region
blue-ward of the Lyman limit. Problems arise when the observed QSO
presents a large individual free path. This effect causes a bluer $u-g$
color, which eventually could be lower than the SDSS threshold used to
select QSOs at $z>3.6$. Fig. \ref{ugMFP} shows the effect discussed
here. In this plot, our QSO sample has been split into two parts using a
threshold in IFP of 150 pMpc, in the cases of sources with higher and lower redshift w.r.t. the median one of the sample ($z\sim3.90$). The $u-g$ distribution with larger
(smaller) free paths with respect to the threshold is represented
by the green (orange hatched) histogram, and its mean value is marked
by the dot-dashed (solid) black line. As can be seen, at larger values
of the IFP, the mean value of the distribution moves to bluer
$u-g$ colors (green histograms), approaching the 1.5 threshold adopted by SDSS to select the
observed QSOs. In particular, this effect is more pronounced at lower redshift (bottom panel) where a possible lack of sources with high IFP would change the evolution of this parameter with redshift. For these reasons, QSOs with large IFP could be lost in the
standard selection procedure adopted by SDSS, perhaps artificially
decreasing the average value and scatter of the free path distribution and
thus the overall contribution of these sources to the UV ionizing
photon production. This effect is particularly severe at $z\lesssim 3.5$,
as been pointed out by \citet{prochaska09} and \citet{cristiani16}.

It is useful to note here that the few QSOs at $3.6\le z\le 4.6$, with
$u-g\le 1.5$ shown in Fig. \ref{ugMFP} (123 out of 2508 sources in the all redshift range), have been observed by SDSS
thanks to other selection criteria, complementary to the optical color
selection of SDSS, as described by \cite{dawson13,paris18}. It is thus not possible to
estimate the systematic effects due to this color selection criterion
with a simple analytic calculation. Just as an example, taking into
account the 16 faint AGNs found at $3.6\le z\le 4.2$ in the COSMOS
field by \cite{boutsia18}, only 5 (approximately 30\%)
would be selected adopting the color
selection of SDSS, with an incompleteness of $\sim$70\%.
Interestingly, the missed AGNs are selected mainly by X-ray emission
\citep{civano16,marchesi16} and show large values of LyC escape
fraction \citep{grazian18} and individual free path. A dedicated survey,
unbiased towards high IFP sources, is required in the future to solve
this issue, which could be particularly severe, both for the proper
estimate of the average value of the free path (and its scatter) and for a
correct quantification of the number density of QSOs at high redshifts.
This survey, however, will have the severe issue of strong contamination by
nearby galaxies and stars, which are affecting the samples
selected by adopting bluer $u-g$ colors.

\subsection{The population of $z\sim 4$ QSOs with low escape fractions and
short free paths}\label{subsec:absorbers}

In the previous sections we have chosen a threshold above 10 pMpc for
the individual free path in order to study the properties of the IGM. We
suspect indeed that the QSOs with a free path lower than this
limit could be affected by associated (intrinsic) absorbers.
At this aim, it is worth noting here that
we do not claim that the
associated absorbers are acting on scales of 10 pMpc. Our hypothesis
is that associated absorbers residing few kpc away from the QSO can have
large peculiar velocities (outflows up to $\sim 10^4$ km s$^{-1}$) and thus they
can mimic a cosmological distance of $\lesssim$10 pMpc from the QSO itself.

With the aim of verifying this guess, we have cross-correlated our
sample with the XQ-100 sample \citep{xq100}, which is a legacy survey
of one hundred quasars at $3.5<z<4.5$ observed with VLT/X-shooter for
a deep exposure time. We find 44 objects in common, and we then refer
to \citet{perrotta16} in order to study the properties of their
absorbers. The latter find that there is a statistical significant (up
to 8$\sigma$) excess of associated NV absorbers at a velocity of $|v|\le
10^4$ km s$^{-1}$, corresponding to $\sim$ 20 pMpc. This is slightly
different with respect to our adopted threshold of 10 pMpc, but it can
be comparable to it, considering the redshift uncertainties of $\sim
0.005$ for the SDSS sample and the intrinsic uncertainties on the IFP
determination of 2-3 pMpc, due to our procedure.

Out of the 44 QSOs in common between our sample and the XQ-100 one, we
check in detail the absorbers of the 6 QSOs with free path less than
10 pMpc. We find that, for all six, the $C_{IV}$ absorbers studied by
\citet{perrotta16} lie within 1-2 pMpc from our estimates, which
indicates these absorbers are the responsible of the short free
path measured by our methods, and of the low value of the escape
fraction, as we have discussed above. More precisely, 2 out of 6 QSOs
have a NV system, which according to \citet{perrotta16} is a robust
signature of associated absorbers. Indeed, our recovered NV fraction
(33\%) is similar to the ones found by \citet{perrotta16} and
\citet{perrotta18}, i.e., 33\% and 38\%, respectively.
The other four systems have no individual detection of NV in
\citet{perrotta16}; however, a more recent work by \citet{perrotta18}
indicates that $C_{IV}$ absorptions with $N(C_{IV})\ge 10^{14} 
$cm$^{-2}$ within 5000 km s$^{-1}$ of the emission redshift of the QSO
(mimicking a distance of $\sim 10$ pMpc) are detected in NV at
15$\sigma$ through stacking, supporting our guess that probably all
these systems are bona-fide associated, given their large column
densities and high excitations. High ionization absorption lines
indeed require close proximity to the QSOs based on detailed
photo-ionization constraints \citep{chen18}. 


\section{Summary and Conclusions}

The aim of this paper is to measure the LyC escape fraction and the free path of HI ionizing photons for a complete sample of the $z\sim 4$ QSO population, as a follow-up study of the work by \citet{cristiani16}. The $f_{esc}$ and MFP are fundamental parameters involved in the characterization of the IGM and, in particular, in the estimate of the HI photo-ionization rate $\Gamma_\mathrm{HI}$ at high redshifts.
At this aim, we select 2508 QSOs at $3.6 \leq z \leq 4.6$ and in the magnitude range $17.0 \leq I \leq 20.0$ ($-29.0 \lesssim M_{1450} \lesssim -26.0$).

In order to quantify the amount of UV photons able to ionize the surrounding IGM, the LyC escape fraction of each QSO was computed as the ratio between the average flux blue-ward ($892 \leq \lambda_{rest} \leq 905$ {\AA}) and red-ward ($915 \leq \lambda_{rest} \leq 960$ {\AA}) of the Lyman limit.
The PDF$(f_{esc})$ resulted in a multi-modal distribution with a mean value $f_{esc} = 0.49 \pm 0.36$. In particular, the narrow peak at $f_{esc} \leq 0.2$ could depend on the wavelength coverage used to compute the average flux in the ionizing region, blue-ward of the Lyman limit. In fact, a shorter spectral window, closer to 912 {\AA} rest-frame, enhances the number of sources with high escape fraction, reducing the height of the peak at low $f_{esc}$ (Fig. \ref{fig:MFP_distr}). This, in turn, increases the average value of the PDF($f_{esc}$), making the result obtained in this paper just a lower limit to the real value of the $f_{esc}$.
Indeed, the mean $f_{esc}$ we find ($\sim0.5$) is consistent with the average flux decrement between 930 and 900 {\AA} rest-frame due to
intervening IGM absorption derived by \cite{inoue14} at $z=4$ ($\sim0.6$, see their Fig. 4), suggesting that the escape fraction could likely reach 100$\%$ in bright QSOs, if we correct for this effect.
This last consideration is then strengthened by the connection between $f_{esc}$ and the individual free path of HI ionizing photons we find in this paper. Sources with IFP $\leq 10$ pMpc also show a low LyC escape fraction (Fig. \ref{fescMFP}); as reported in Section \ref{sec:MFP_vs_esc}, this is again an effect of the spectral window chosen to compute $f_{esc}$.
Furthermore, QSOs with IFP $\leq 10$ pMpc
(possibly affected by associated absorbers)
should not be ideal tracers of 
the properties of the IGM.
Finally, a redshift and magnitude evolution of $f_{esc}$ was excluded.
Therefore, the ionizing properties of faint AGNs and bright QSOs seem to be the same at $z \sim 4$, in agreement with \citet{grazian18}.

Regarding the estimate of the mean free path of HI ionizing photons from its distribution function, this is a method which could represent an improvement with respect to the analysis performed so far in the literature. Indeed, the classical approach to the calculation of the mean free path relies on assumptions on the frequency distribution of IGM absorbers or on the direct estimate of this parameter from the stacked spectra of QSOs in different redshift bins. On one side, the first approach is uncertain because of several issues on the derivation of the exact HI frequency distribution; on the other side, the stack technique seems to depend on the fraction of QSOs with low individual free paths.
For this reason, the statistical analysis of the IFP distribution has been considered a more informative method to study the ionization status of the IGM.

Summarizing, we have found that:
1-the IFP distribution function carries more information than the MFP from stacking;
2-the MFP at $z\sim 4$ are larger than the one found by
\citet{worseck14} by a factor of 1.1-1.7, probably due to their sampling fainter luminosities at higher redshifts and not cleaning their sample for possible associated (intrinsic) absorbers; 3-the redshift evolution of the MFP we derive is milder
than the one obtained by \citet{worseck14}; 4-if the trends found at $z\sim 4$ and $M_{1450}\sim -27$ are confirmed also at higher redshifts and at fainter luminosities, this could have strong implications for the role of bright QSOs and faint AGNs on the hydrogen reionization process.

As we did for the LyC escape fraction, we check the dependence of the free path on luminosity and redshift. For the first case, we divide the sample in redshift bins of $\Delta z = 0.2$, each of which has been in turn split in magnitude bins $\Delta M_{1450}=1.0$. The IFP of bright QSOs turns out to be
larger than the one of faint QSOs, at least up to $z=4$. This trend becomes milder going at higher redshifts and seems not to be affected by systematic effects.

As previously said about the free path, we used the same redshift and absolute magnitude bins to test its redshift evolution by computing the MFPs from the PDF in each bin; then, we compared our results with the ones of \citet{worseck14} w.r.t. which we found a milder trend of MFP versus redshift. This is mainly due to the technique involved in the calculation of the MFP by \citet{worseck14}, i.e., the stack method, which could give an underestimate of the real value of the MFP at different redshifts (see Appendix \ref{appendix:B}). Excluding QSOs with IFP $\le 10$ pMpc, assuming that these sources are affected by associated absorbers, makes the difference between the MFPs from the PDFs and the ones obtained by the stack even larger (MFP = 49 - 59 pMpc at $z = 4$, from our work, w.r.t. MFP = $37 \pm 2$ pMpc, by \citet{worseck14}), with milder evolution with redshift ($\eta \sim -2.6 \div -4.3$ versus the best fit of \citet{worseck14}, i.e., $\eta = -5.4$).
Extrapolating these results at higher redshifts could have important implications about the reionization process.

In this regard, we have computed the scaling factor $\Gamma_\mathrm{HI}^\mathrm{new}/\Gamma_\mathrm{HI}^\mathrm{old}$ in order to correct previous estimates of the photo-ionization rate according to our new measurements of MFPs from the PDFs.
To do that, we consider two extreme cases:
1-all the QSOs with IFP $\le 10$ pMpc were affected by intervening absorbers ("option 1"); 2-that these latter sources were all associated (intrinsic) systems which should be rejected for a correct estimates of the mean free path ("option 2").
In both cases, for an ionizing background mainly produced by QSOs and AGNs, the correction factor goes from 1.2 at z=3.7 to 1.7 at z=4.5, respectively, with a possible monotonic trend with redshift.
It is worth noting that recent results of \citet{perrotta16,perrotta18} indicate that at least 30\%
of the CIV absorbers at $|v|\le 10^4$ km s$^{-1}$ (i.e., at $\le 10-20$ pMpc if we translate the velocity offset into a cosmological distance) from the QSOs could be affected by associated (intrinsic) systems.

Finally, it is worth noting that our estimate of the PDF of the free path
at $z\ge 3.5$ is of paramount importance since it gives a reference
benchmark for the theoretical models which are investigating the
reionization epoch with state-of-the-art cosmological hydrodynamical
simulations with complex radiative transfer calculations. It will be
interesting to check whether these models are able to reproduce the
mean value, the scatter, and the redshift evolution of the IFP distribution functions
outlined in this paper.
Detailed comparison with state-of-the-art simulations,
as the ones carried out by \citet{inoue14} will be the
subject of a future paper.

Although these ones are important clues for studying the evolution of
the MFP with redshift, it is necessary to go deeper in magnitude, by
observing fainter AGNs at $z\ge 4$ in order to increase the statistics on these
sources and thus to better investigate their possible luminosity
evolution. This will be done by a follow-up survey, extending the
seminal work done by \citet{grazian18}. In addition, extending in
redshift at $z\sim 5-6$ the analyzed sample of QSOs is of paramount importance to
confirm the trends we have found at $z\le 4.6$.


\begin{acknowledgements}
We warmly thank the referees for their deep insight into the paper, for the careful reading of our manuscript, for the constructive comments and useful suggestions that contribute to improve the quality of our paper.
Funding for the Sloan Digital Sky Survey IV has been provided by the Alfred P. Sloan Foundation, the U.S. Department of Energy Office of Science, and the Participating Institutions. SDSS-IV acknowledges
support and resources from the Center for High-Performance Computing at
the University of Utah. The SDSS web site is www.sdss.org.
SDSS-IV is managed by the Astrophysical Research Consortium for the 
Participating Institutions of the SDSS Collaboration including the 
Brazilian Participation Group, the Carnegie Institution for Science, 
Carnegie Mellon University, the Chilean Participation Group, the French Participation Group, Harvard-Smithsonian Center for Astrophysics, 
Instituto de Astrof\'isica de Canarias, The Johns Hopkins University, Kavli Institute for the Physics and Mathematics of the Universe (IPMU) / 
University of Tokyo, the Korean Participation Group, Lawrence Berkeley National Laboratory, 
Leibniz Institut f\"ur Astrophysik Potsdam (AIP),  
Max-Planck-Institut f\"ur Astronomie (MPIA Heidelberg), 
Max-Planck-Institut f\"ur Astrophysik (MPA Garching), 
Max-Planck-Institut f\"ur Extraterrestrische Physik (MPE), 
National Astronomical Observatories of China, New Mexico State University, 
New York University, University of Notre Dame, 
Observat\'ario Nacional / MCTI, The Ohio State University, 
Pennsylvania State University, Shanghai Astronomical Observatory, 
United Kingdom Participation Group,
Universidad Nacional Aut\'onoma de M\'exico, University of Arizona, 
University of Colorado Boulder, University of Oxford, University of Portsmouth, 
University of Utah, University of Virginia, University of Washington, University of Wisconsin, 
Vanderbilt University, and Yale University.
\end{acknowledgements}

\begin{appendix} 

\section{Example of SDSS QSO spectra}

\begin{figure}
\centering
\includegraphics[width=\columnwidth,angle=0]{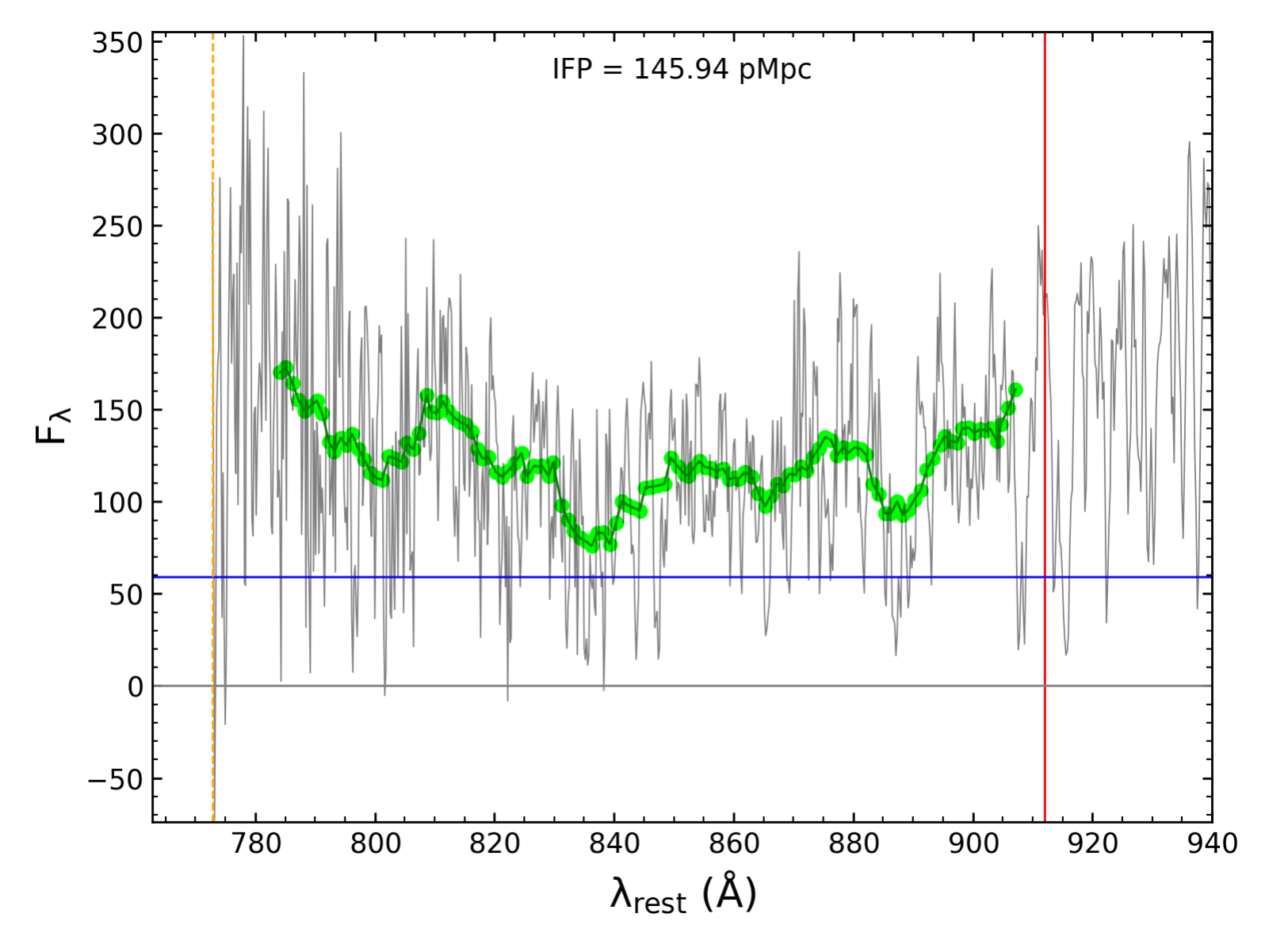}
\caption{
Free path estimate of the QSO SDSS J113654.6+485322.3
at $z_\mathrm{spec}=3.61964$. Same colors legend as in Fig. \ref{fig:MFP4654}. 
}
\label{esMFP}
\end{figure}

In this appendix, we report another example (beyond the one previously shown in Fig. \ref{fig:MFP4654}), which explains the procedure adopted to estimate the IFP for each SDSS QSO. In particular, Fig. \ref{esMFP} is an example of the issue related to the estimate of the free path for several sources in our sample, i.e., the limited wavelength coverage of some SDSS spectra. Here the LyC emission extends at wavelengths bluer than the lower limit of the SDSS spectrum, i.e., $\lambda_{obs}=3560$ {\AA}, indicated
by the dashed orange line, but does not intersect the blue line, which marks the $1/e$ level w.r.t. the flux at 912 {\AA} rest-frame. In this case the estimated IFP of 145.94 pMpc is only a lower limit due to the instrumental
limitations, and the true free path for this QSO could be even larger.

\section{The mean free path from stack spectra and from the mean value of the
probability distribution function}\label{appendix:B}

The results from the statistical analysis on the IFP distribution
functions for the high-z QSOs highlight a possible disagreement
between these latter values and those from the stacks. In principle,
this could be caused by the shape of the individual free path distributions,
which result to be highly asymmetric/skewed and wide at $z\ge 3.6$. To test
this hypothesis, a comparison between the MFP from the stacked spectra
and from the mean value of the distributions of sub-samples selected
in narrow intervals of IFP in the redshift range $3.60\le z\le 3.80$
was accomplished. Table \ref{tabMFPart} summarizes the obtained
results. As reported in the first three rows of this table, bins of 10
pMpc in IFP were considered, namely between 50-60, 90-100, and 140-150
pMpc (column 1); in each of these bins, a number $N_\mathrm{obj}$ of QSOs was
found (column 2). Comparing the mean free path from the stacks and
distributions (columns 3 and 4, respectively), it can be seen that the
discrepancies are clearly reduced.

Furthermore, two distant peaks were also analyzed in order to study
the effect of the asymmetry/skewness of the distribution on the stacked
spectrum. The obtained values are reported in the last row of Table
\ref{tabMFPart} for two peaks at 15 and 105 pMpc, with 36 and 8 QSOs in
each bin, respectively. In this case, the results from the stack and
the distribution are quite different from each other, being the former
value significantly lower than the latter one. These observations
seem to lead to the conclusion that in a wide and really asymmetric
distribution, the stack is dominated by sources with small IFP,
resulting in a lower value w.r.t. the one obtained by the
distribution function. Moreover, the stack technique seems to depend
on the number of sources with low free path with respect to those with
higher values of this parameter. Indeed, if the number of QSOs in the
bin with low free paths is about or lower than twice of that in
the highest bin, the stacked spectrum is dominated by these latter
sources, resulting in the estimate of a large MFP. On the other
hand, if the ratio between the two numbers is major than three (which
is the case discussed here), sources with low free paths dominate
the stack, which provides a result lower than the one computed from
the mean of the associated distribution. This is a simple consequence
of the definition of IFP, which is the distance where the emitted flux
is reduced by a factor of $1/e\sim 1/3$, i.e., thus dominated by the
bulk of the IFP distribution.

Finally, it is reasonable to believe that the disagreement between the
two methods discussed above could increase going at higher redshift
(i.e., $z > 4$) because of the likely more pronounced asymmetry of the
IFP distribution functions. Certainly, approaching the end of the
reionization at $z\sim 6-7$, the discrepancy should be reduced,
because of the decrement of the free paths for all the sources,
resulting in a PDF peaked at low values. 

The simulations carried out here are quite simplistic and it is possible that they are not able to fully reproduce the real IFP distributions. However, they are useful in order to understand the behavior of the stacking procedure when applied to skewed distributions, as the ones shown in Fig. \ref{fig:MFP_distr}.

\begin{table}
\caption{Stack vs Probability Distribution Function from simulations. Comparison between the estimate of the MFP from the stacked spectra
and the mean of the probability distribution functions. Narrow and
distant bins were analyzed in order to
study the effect of the asymmetry and of the width of the
distributions on the stack technique. The studied bins, their number
of sources, and the values obtained from the stacks and the
distributions are reported from the first to the last column. The last
row reports the results of the analysis of two distant bins at 15 and
105 pMpc with 36 and 8 QSOs, respectively, in order to study the effect
of the asymmetry of the distribution on the stacked spectra.}
\label{tabMFPart}
\centering
\begin{tabular}{l c c c}
\hline
\hline
MFP interval & $N_\mathrm{obj}$ & MFP$_\mathrm{stack}$ & MFP$_\mathrm{mean}$ \\
pMpc & & pMpc & pMpc \\
\hline
50-60 & 36 & 54.67 & 54.46 \\
90-100 & 22 & 98.84 & 96.08 \\
140-150 & 11 & 150.82 & 144.66 \\
15/105 & 36/8 & 20.26 & 28.86 \\
\hline
\hline
\end{tabular}

\end{table}

\end{appendix}

\end{document}